\documentclass{aastex631}

\usepackage{subfigure}
\usepackage[misc]{ifsym}
\shorttitle{Astrometric observations of NEO using image fusion technique}
\shortauthors{Zhang et al.}
\graphicspath{{./}{figures/}}
\begin{document}
\bibliographystyle{aasjournal}
\title{Astrometric observations of a near-Earth object using the image fusion technique\footnote{Released on 2021 October 29, and this manuscript has been accepted by AJ.}}
\author[0000-0003-3657-5108]{Yigong Zhang$^{(\textrm{\Letter})}$}
\email{zhyg@ynao.ac.cn}
\affiliation{Yunnan Astronomical Observatories, Chinese Academy of Sciences, Kunming 650216, China}
\affiliation{Key Laboratory for the Structure and Evolution of Celestial Objects, Chinese Academy of Sciences, Kunming 650216, China}
\affiliation{Center for Astronomical Mega-Science, Chinese Academy of Sciences, Beijing 100012, China}
\affiliation{University of Chinese Academy of Sciences, Beijing 100049, China}

\author{Jiancheng Wang$^{(\textrm{\Letter})}$}
\email{jcwang@ynao.ac.cn}
\affiliation{Yunnan Astronomical Observatories, Chinese Academy of Sciences, Kunming 650216, China}
\affiliation{Key Laboratory for the Structure and Evolution of Celestial Objects, Chinese Academy of Sciences, Kunming 650216, China}
\affiliation{Center for Astronomical Mega-Science, Chinese Academy of Sciences, Beijing 100012, China}
\affiliation{University of Chinese Academy of Sciences, Beijing 100049, China}

\author{ Jie Su}
\affiliation{Yunnan Astronomical Observatories, Chinese Academy of Sciences, Kunming 650216, China}
\affiliation{Key Laboratory for the Structure and Evolution of Celestial Objects, Chinese Academy of Sciences, Kunming 650216, China}
\affiliation{Center for Astronomical Mega-Science, Chinese Academy of Sciences, Beijing 100012, China}
\affiliation{University of Chinese Academy of Sciences, Beijing 100049, China}

\author{Xiangming Cheng}
\affiliation{Yunnan Astronomical Observatories, Chinese Academy of Sciences, Kunming 650216, China}
\affiliation{Key Laboratory for the Structure and Evolution of Celestial Objects, Chinese Academy of Sciences, Kunming 650216, China}
\affiliation{Center for Astronomical Mega-Science, Chinese Academy of Sciences, Beijing 100012, China}
\affiliation{University of Chinese Academy of Sciences, Beijing 100049, China}

\author{ Zhenjun Zhang}
\affiliation{Anyang institute of technology, Anyang 455000, China}

\begin{abstract}

The precise astrometric observation of small near-Earth objects (NEOs) is an important observational research topic in the astrometric discipline, which greatly promotes multidisciplinary research, such as the origin and evolution of the solar system, the detection and early warning of small NEOs, and deep-space navigation. The characteristics of small NEOs, such as faintness and fast moving speed, restrict the accuracy and precision of their astrometric observations. In the paper, we present a method to improve the accurate and precise astrometric positions of NEOs based on image fusion technique. The noise analysis and astrometric test from the observed images of the open cluster M23 are given. Using the image fusion technique, we obtain the sets of superimposed images and original images containing reference stars and moving targets respectively. The final fused image set includes background stars with high signal-to-noise ratios and ideal NEO images simultaneously and avoids the saturation of background stars. Using the fused images, we can reduce the influence of telescope tracking and NEO ephemeris errors on astrometric observations, and our results indicate that the accuracy and precision of NEO Eros astrometry are improved obviously after we choose suitable image fuse mode.

\end{abstract}


\section{Introduction} \label{sec:intro}

Astrometric observations of small celestial bodies in the solar system are of great significance for the study of the Solar System Formation and Evolution \citep{gomes2005origin, minton2010dynamical, demeo2014solar}, the development of ephemerides for Solar System Objects by follow-up observations \citep{thuillot2011espace, xi2015astrometry, zhang2019ccd} and deep-space exploration \citep{meurisse2020past, buczkowski2020tectonic}. As a part of the solar system, the NEOs have attracted much attention because of their posing threats to the earth and human civilization \citep{bancelin2012asteroid}.  Incidents of small NEOs hitting the earth have occurred frequently, such as the ``Tunguska Event'' occurred on 1908 June 30 in the Tunguska region of Siberia \citep{ganapathy1983tunguska, 2014The}, the meteorite impact event occurred in Chelyabinsk, Russia on 2013 February 15 \citep{brown2013500}, and the bolide event occurred in Diqing, Yunnan Province on 2017 October 4 \citep{ye2018preliminary}. The precise astrometric observations of small objects in the solar system, especially small NEOs, will help us to monitor their orbits for avoiding the potential threats to the Earth.

Using the traditional long-exposure methods to observe fast-moving faint and small objects, such as fixed star tacking mode or moving object tracking mode \citep{ryan2008magdalena, zhang2019ccd}, we obtain the streaked images of targets that affect the accuracy of the astrometry. Many method are available to avoid streaks in a single image, such as the ``shift-and-stack'' technique for tracking Kuiper Belt objects \citep{tyson1992limits, bernstein2004size}, and the synthetic tracking technique for searching NEOs \citep{shao2014finding, zhai2014detection, zhai2018accurate, zhai2020synthetic}. Image fusion is to superimpose multiple images into an image, in which the information complementarity, temporal, and spatial correlation in different images are used, the superimposed image contains more comprehensive and detailed information than original images. For a set of images with background stars and moving objects, the streaks will appear when the relative velocity between stars and moving objects with respect to exposure time is too fast, and the streaks still remain even if the images are obtained using the ``shift-and-stack'' method. If we take a set of images containing background stars and moving objects, and for the background stars and moving objects using the ``shift-and-stack'' method, respectively. We can obtain a high-quality image, which contains many background stars and moving objects with high signal-to-noise ratios (S/Ns), using an image fusion technique, in which the saturation of brighter background stars can be avoided.

The outline of the paper is as follows. In Section 2, the noise analysis and astrometric test on the superimposed image of the open cluster M23 are presented in detail. The methods to realize astrometry using image fusion technique are given in Section 3. In Section 4, we present a solution to get the observed images of NEOs, and introduce the limits of applicability of the image fusion technique. Section 5 shows the astrometric results of near-Earth asteroid Eros using different methods for comparison. Finally, some conclusions are shown in Section 6.

\section{Noise analysis and astrometric test of the superimposed image} \label{sec:noise}

The noise in the observed image mainly contains two parts: the poisson noise of objects and background noise in the image. The related work of astrometric observations in this paper is mainly involves the targets that can be detected in a single image; therefore, the poisson noise of target is not considered in this paper, and we focus on the analysis of the background noise in the observed image. Noise analysis on background noise mainly includes dark current noise ${N}_{dark}$, skylight background ${N}_{sky}$ and readout noise $ {N}_{readout}$ \citep{sun2012new}. The SNR can be expressed by
\begin{equation}
\frac{S}{N} = \frac{S}{\sqrt{S+{N}_{sky}+{N}_{readout}^2+{N}_{dark}^2}},
\end{equation}
where $S$ is the signal of target source, and ${N}$ is the noise.

There are two cases for noise analysis.
	
Case one: for a single image with a long exposure time and a superimposed image with long total exposure time stacked by $n$ images, if the exposure time of individual image using for stacking is long enough, the photon signal will be much higher than the read noise and dark current noise which can be ignored. When the total exposure time of the superimposed image is equal to that of the long-exposure image, and the S/N of the long exposure image or superimposed image is given by
\begin{equation}
\frac{S_{Long}}{{N}_{long}} \approx \frac{S_{stack}}{{N}_{stack}}\approx\frac{n\times S_{single}}{\sqrt{n\times S_{single}+n\times {N}_{sky}}}\approx\sqrt{n}\times \frac{S_{single}}{{N}_{single}},
\end{equation}
where $S_{long}$ is the signal of target in image with long exposure time, ${N}_{long}$ is the noise in image with long exposure time, $S_{stack}$ is the signal of target in superimposed image, ${N}_{stack}$ is the noise in superimposed image, $S_{single}$ is the signal of target in single image, and ${N}_{single}$ is the noise in single image. In this case, the S/N of the superimposed image is increased approximately by the factor of $\sqrt{n}$ compared with single image, and approximately equals to that of the image with long exposure time.

Case two: if the exposure time of individual image using for stacking is short enough, the readout noise will be dominant, and the photon signal and dark current noise can be ignored. In this case, the S/Ns of the long exposure image and superimposed image are given by 
\begin{equation}
\frac{S_{long}}{{N}_{long}} \approx \frac{n \times S_{single}}{{N}_{readout}},
\end{equation}

\begin{equation}
\frac{S_{stack}}{{N}_{stack}}\approx \frac{n\times S_{single}}{\sqrt{n}\times{N}_{readout}}\approx\sqrt{n}\times\frac {S_{single}}{N_{single}}\approx \frac{1}{\sqrt{n}}\times\frac{S_{long}}{{N}_{long}},
\end{equation}
showing that the S/N of superimposed image is also approximately $\sqrt{n}$ (the increase rate is not always approximately $\sqrt{n}$, it mainly depends on the specifications of detector and the length of exposure time) times larger than that of the single image as same as case one, but it is approximately $\frac{1}{\sqrt{n}}$ times compared to that of the image with long exposure time. We know that image stacking processing can improve the S/N of image in two cases. 

On the evening of 2014 May 29 we used the 1-m optical telescope of the Yunnan Observatory to observe the open cluster M23 and obtained a data set. There are seven images in this data set. The details about the telescope and CCD detector are listed in Table 1. The site (i.e., IAU code 286) is located at longitude E102.788°, latitude N25.0294°, and height 2000 m above sea level.
\begin{deluxetable*}{cc}
	\tablenum{1}
	\centering
	\tablecaption{Specifications of telescope and CCD Detector\label{tab:specification}}
	\tablewidth{10pt}
	\tablehead{\colhead{Description} & \colhead{Value}}
	\startdata
	Approximate focal length & 1330cm\\
	F-ratio & 13\\
	Diameter of primary mirror & 101.6cm\\
	CCD filed of view & $7.1'\times 7.1'$\\
	Size of pixel & $2048 \times 2048$\\
	Approximate angular extent per pixel & $0.21''$\\
	\enddata
\end{deluxetable*}

\begin{figure}[ht]
	\centering
	\subfigure{\includegraphics[scale=0.5]{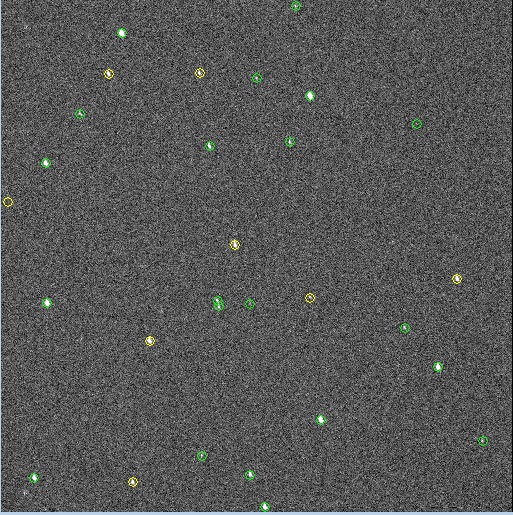}}
	\subfigure{\includegraphics[scale=0.5]{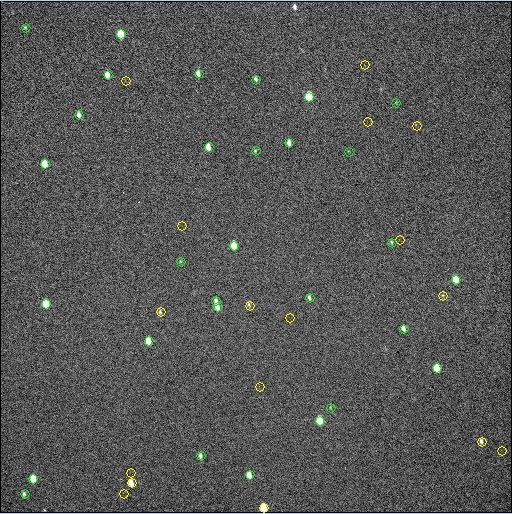}}
	\caption{The original and superimposed images. The left image is the original image with an exposure time of 60 ms, not using filter and observed on 2014 May 29. The right image is the superimposed image using seven original images, where the total exposure time is 420ms.  The green circles represent the stars with smaller residual comparing with the stars marked by yellow circles.  }
	\label{fig:fig1}
\end{figure}

The left-hand panel of Figure 1 shows an original image. We align and stack all original images by using an original image as reference image. There are seven images in the original data set, therefore a data set containing seven superimposed images is obtained. The right-hand panel of Figure 1 shows a superimposed image. The original and superimposed image sets are reduced by astrometry,  respectively. The residuals in R.A. (R.A.) and decl. (decl.) are shown in Figure 2. The results indicate that the number of reference stars in the superimposed image has increased significantly than that of the original image, and the precision of astrometric observations is significantly improved.

We use the interactive software tool Astrometrica as the astrometric solution (\break\url{http://www.astrometrica.at/}). There are 29 stars and 47 stars found in the original and superimposed images,  respectively. The reference stars used for the astrometric data reduction are shown by green circles. 'Bad' reference stars (their residuals in either R.A. or decl. are larger than $0.2''$) are marked by yellow circles. Finally there are 21 stars and 30 stars used for the astrometric data reduction in the original and superimposed image respectively.

\begin{figure}[ht]
	\centering
	\subfigure{\includegraphics[scale=0.31]{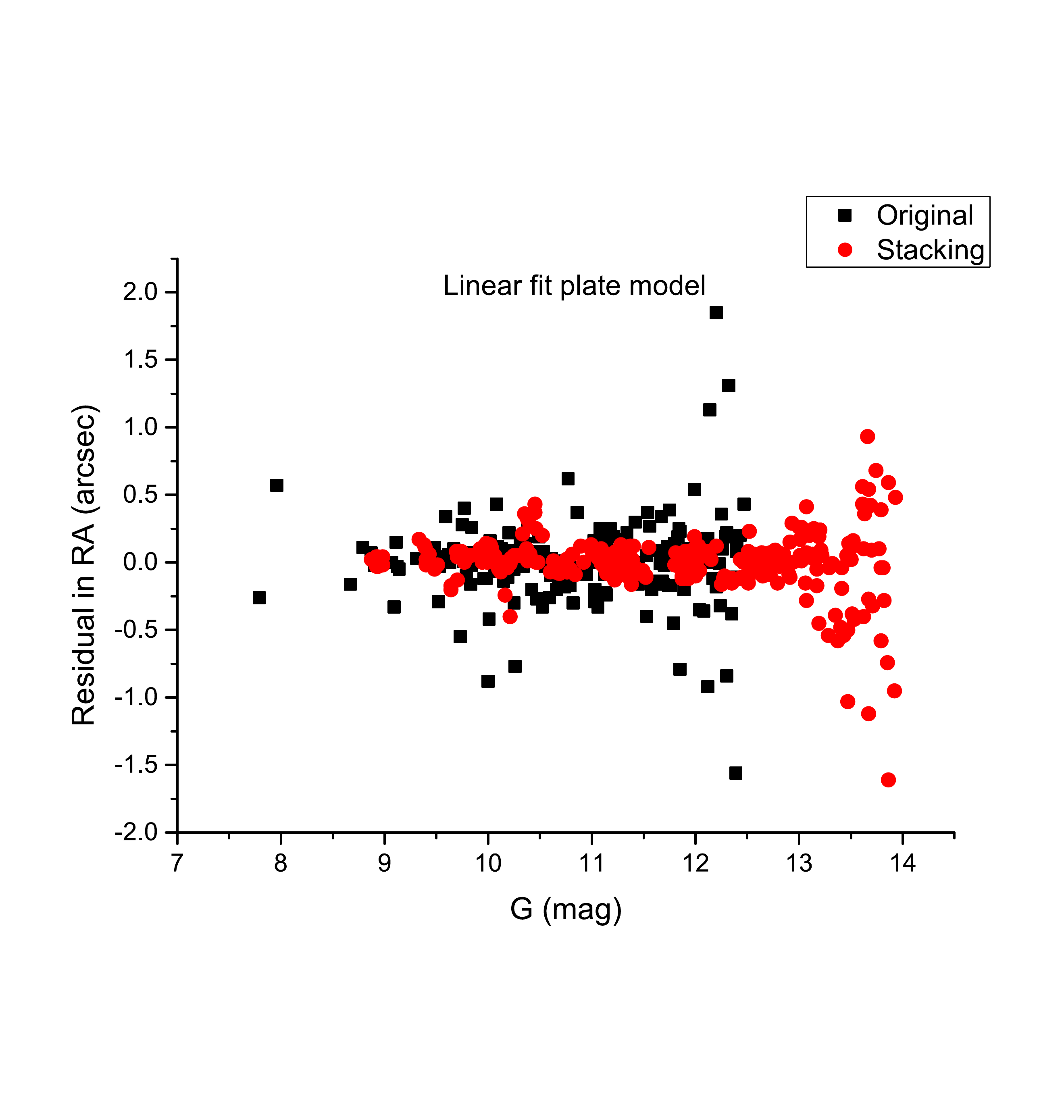}}
	\subfigure{\includegraphics[scale=0.31]{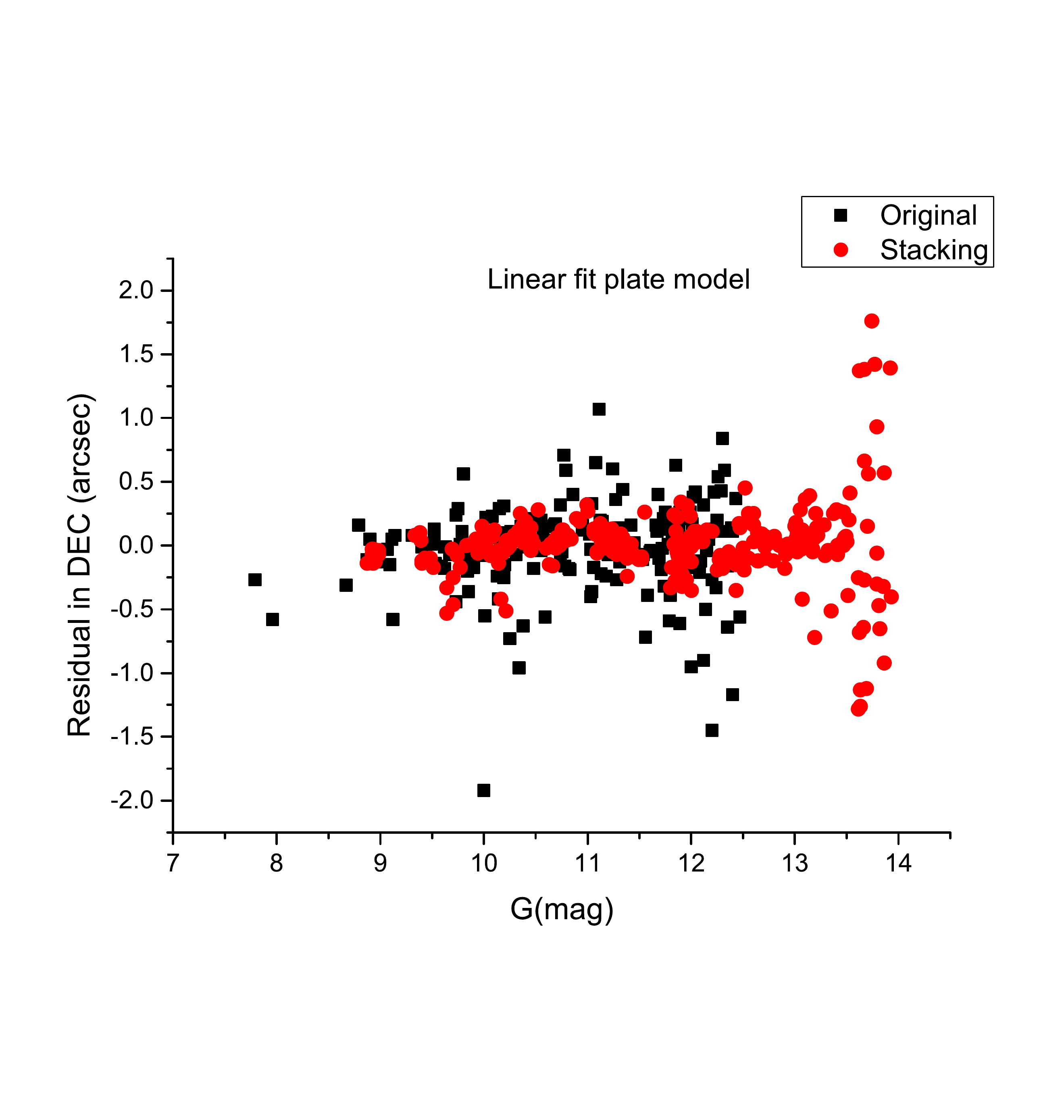}}
	\subfigure{\includegraphics[scale=0.31]{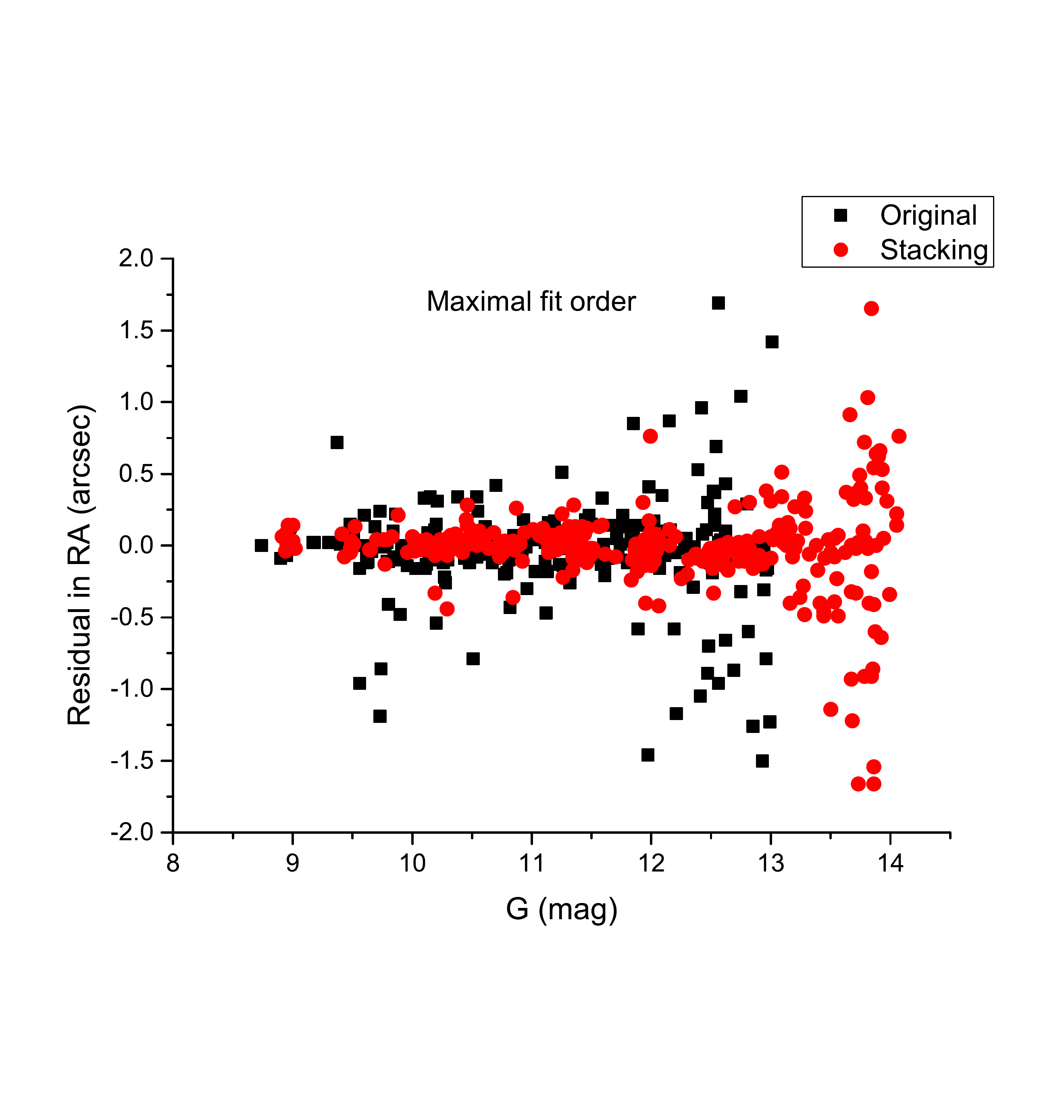}}
	\subfigure{\includegraphics[scale=0.31]{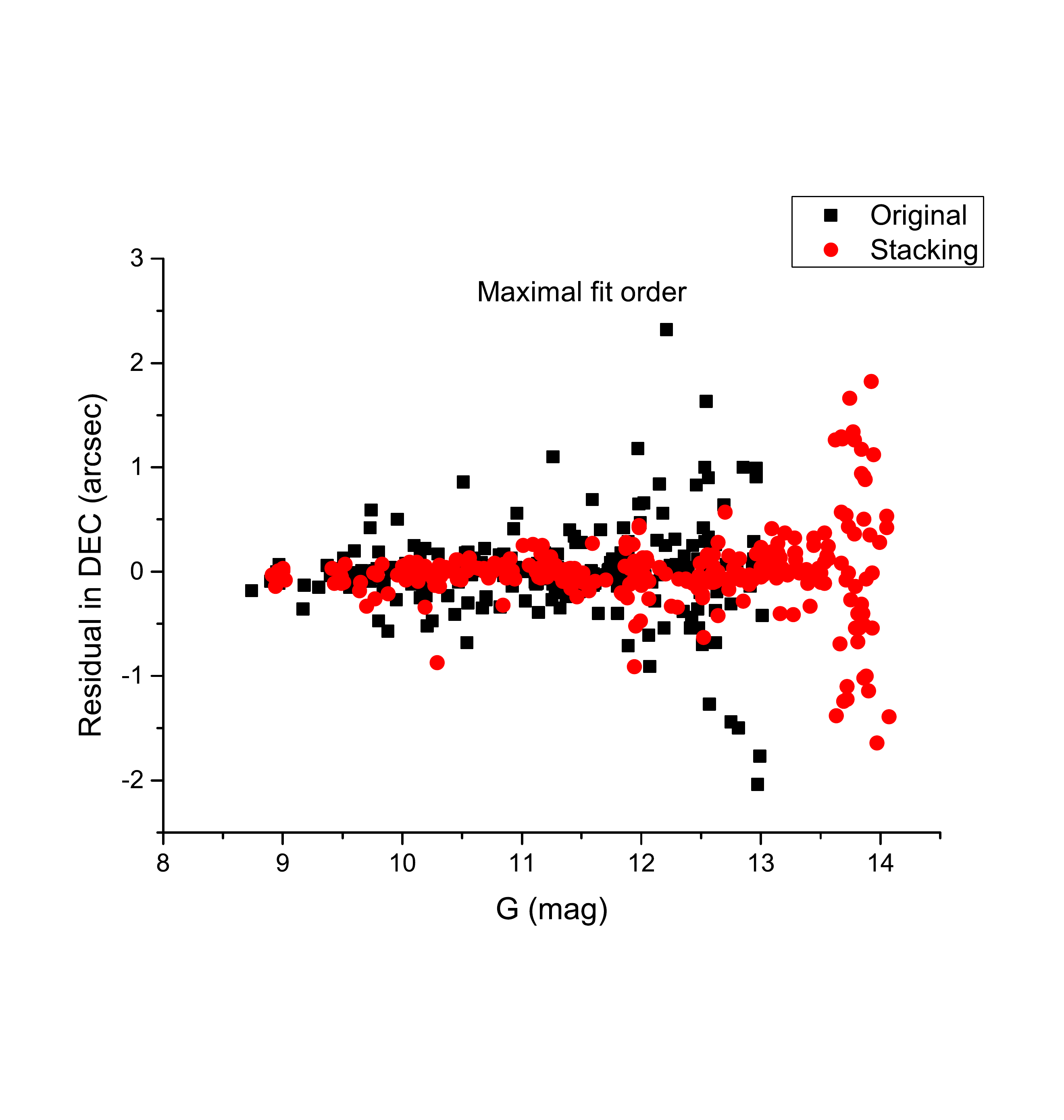}}
	\caption{The residuals of open cluster M23 in R.A. and decl. with different plate models. The black solid squares signify the residuals based on original image set and the red points represent the residuals based on superimposed image set. Obviously, the dispersion of residuals in R.A. and decl. obtained using superimposed image set is smaller than using original image set. }
	\label{fig:fig2}
\end{figure}

In astrometric reduction, the choice of the plate constant model is very important. Usually we need about $\frac{n^2}{4} $ reference stars to prevent the overfitting of plate constants, where n represents the number of plate constants. For example, at least nine stars are required to calculate the plate constants for the six-constant model. Figure 1 shows the star images of open cluster M23 after superposition processing, the number of detected reference stars has increased from 29 to 47. The number of reference stars used for the astrometric data reduction has increased from 21 to 30 in the superimposed image. To show the improvement of astrometry in superimposed image, we use linear and maximal fit models (when the reference stars do not satisfy the initial cubic fit, the software will automatically switch to lower fit) to solve the plate constants, respectively. The results are shown in Figure 2, indicating that the magnitude limits of detected reference stars increase in superimposed image compared to that in original image and the astrometric residuals of superimposed images are also smaller than that of original images no matter of plate models used. More reference stars are benefitted for the reliable determination of plate constants. The results of \citet{lin2019characterization} indicate that the brighter reference star within $100''$ from the target can significantly increase the astrometric accuracy of the target in the field of view. Therefore, the improvement for the accuracy and precision of astrometric observations by increasing the number of reference stars in the field of view is not only applicable to small field telescopes, but also to large area telescopes.

\section{Methods} \label{sec:method}
We introduce the method of the image fusion technique. The main idea is that the multiple images of reference stars and moving objects are independently fused to form a fused image set. The images in the fused image set contain more reference stars with high S/N and a high-quality image of the moving target. The image fusion has the following steps.

Step 1. We select a near-Earth asteroid as the target to analyze and determine the exposure time of a single image, we then take observations to get a set of images.

For the selected target, the exposure time of single observation depends on the apparent magnitude of the target, the parameters of the telescope and CCD detector, and the skylight background. The S/N and possible tailing of the target
are also considered. In this paper, we use a criterion to confirm the quality of target image given by
\begin{equation}
Flag=({SNR}_{Asteroid}\geq 3)\&\&({FWHM}_{Asteroid}\textless 1.1\times{FWHM}_{star}),
\end{equation}
where $Flag$ is an identifier, ${S/N}_{Asteroid}$ is the S/N of the target, ${FWHM}_{Asteroid}$  is the half maximum full width of the target, and ${FWHM}_{star}$  is the half maximum full width of a brighter and unsaturated star in the field of view.

When Flag = 0, it means that the exposure time of the target needs to be increased.

When Flag = 1, it means that the target has a good image with a suitable exposure time, a satisfactory S/N and no tailing.

With the above steps, the multiple images are taken as a data set. In the paper, we select the asteroid 00433 named Eros to describe the method.
The image sets are observed with the 1 m telescope of the Yunnan Observatory in 2019. One image of Eros	is shown in Figure 3.

\begin{figure}[ht]
	\centering
	\includegraphics[scale=0.8]{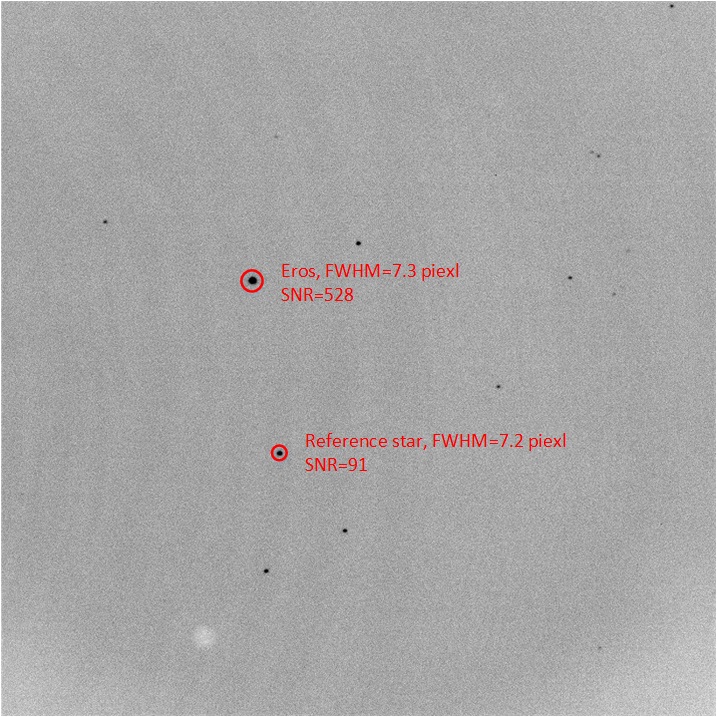}
	\caption{The original image of near-Earth asteroid Eros. The image is obtained by \textit{I} band observation, and the exposure time is 15 seconds.
The FWHM and S/N satisfy the criterion of equation (5).}
	\label{fig:fig3}
\end{figure}

Step 2. Image preprocessing.

Image preprocessing includes the calibrations of flat field, bias, and dark, shown by
\begin{equation}
{S}^{*} = \frac{S-{S}_{dark}-{S}_{bias}}{{S}_{flat}-{S}_{dark}-{S}_{bias}},
\end{equation}
where ${S}^{*}$ is the preprocessed data, $S$ is the observed data, ${S}_{dark}$ is the dark data, ${S}_{flat}$ is the flat field data, and ${S}_{bias}$ is the bias data.

Step 3. We divide the reference stars and near-Earth asteroids into two data sets, in which the asteroid region is segmented from the original image and the remaining region only contain reference stars. The data of image segmentation can be described by
\begin{equation}
{Data}_{Asteroid} + {Data}_{Star}={Data}_{Full},
\end{equation}
where ${Data}_{Full}$ is the preprocessed original image data, ${Data}_{Asteroid}$ is the observed asteroid image data after segmentation, and ${Data}_{Star}$ is the segmented reference star data. ${Data}_{Asteroid}$ is extracted according to the fixed matrix size, in which the size of extraction area mainly depends on the velocity of moving target and the time span of original data set, and the data of other region assigned to be zero. The reference star data is given by
\begin{equation}
{Data}_{Star}={Data}_{Full}-{Data}_{Asteroid}.
\end{equation}
Both ${Data}_{Asteroid}$ and ${Data}_{Star}$ are saved in the same format as the original image. Figure 4 shows the processing result of this step.

\begin{figure}[ht]
	\centering
	\subfigure{\includegraphics[scale=0.2]{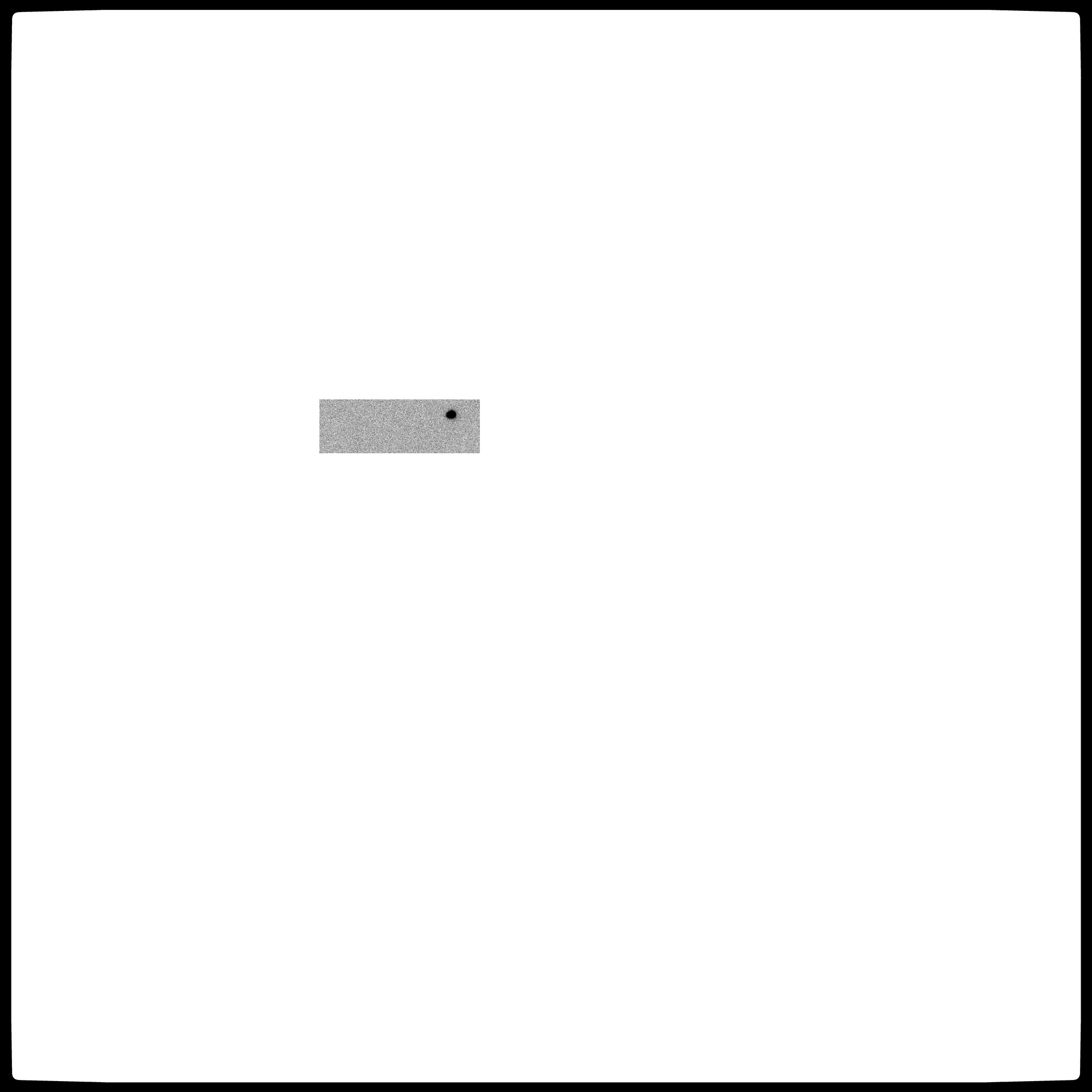}}
	\subfigure{\includegraphics[scale=0.2]{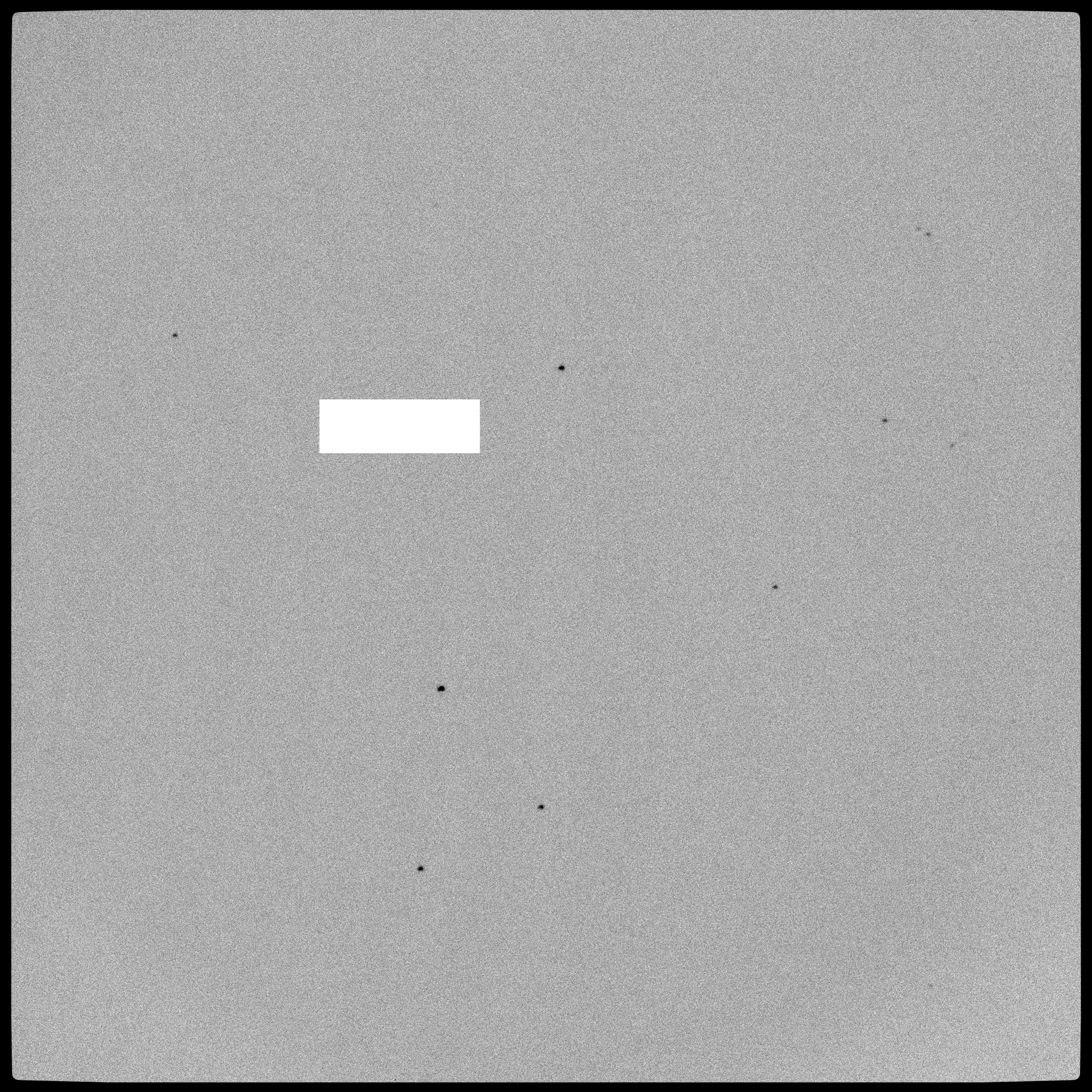}}
	\caption{The segmented asteroid and reference star images. The left image is the segmented asteroid, and the right image is the segmented reference stars. }
	\label{fig:fig4}
\end{figure}

Step 4. We select a reference star image as the reference image. Aligning and stacking other images with this image, we get the superimposed reference image. The calculation method is presented as follows. First, we calculate the center coordinates of stars in the reference image and other images, defined as ${P}_{r}(x_r,y_r )$ and ${P}_{n}(x_n,y_n )$ , respectively. The relation between ${P}_{r}(x_r,y_r )$ and ${P}_{n}(x_n,y_n )$ is given by
\begin{equation}
{P}_{r}(x_r,y_r ) \stackrel{A}{\longleftrightarrow}{P}_{n}(x_n,y_n ),
\end{equation}
where $A$ is the affine transformation matrix which can be obtained by using a known data of ${P}_{r}(x_r,y_r )$ and ${P}_{n}(x_n,y_n )$. We then align and stack other images with the reference image based on affine transformation matrix A. The gray-scale value of the superimposed image is the mean gray-scale value of original images in the correspondence regions. Figure 5 shows the processing result of this step.

\begin{figure}[p]
	\centering
	\includegraphics[scale=0.25]{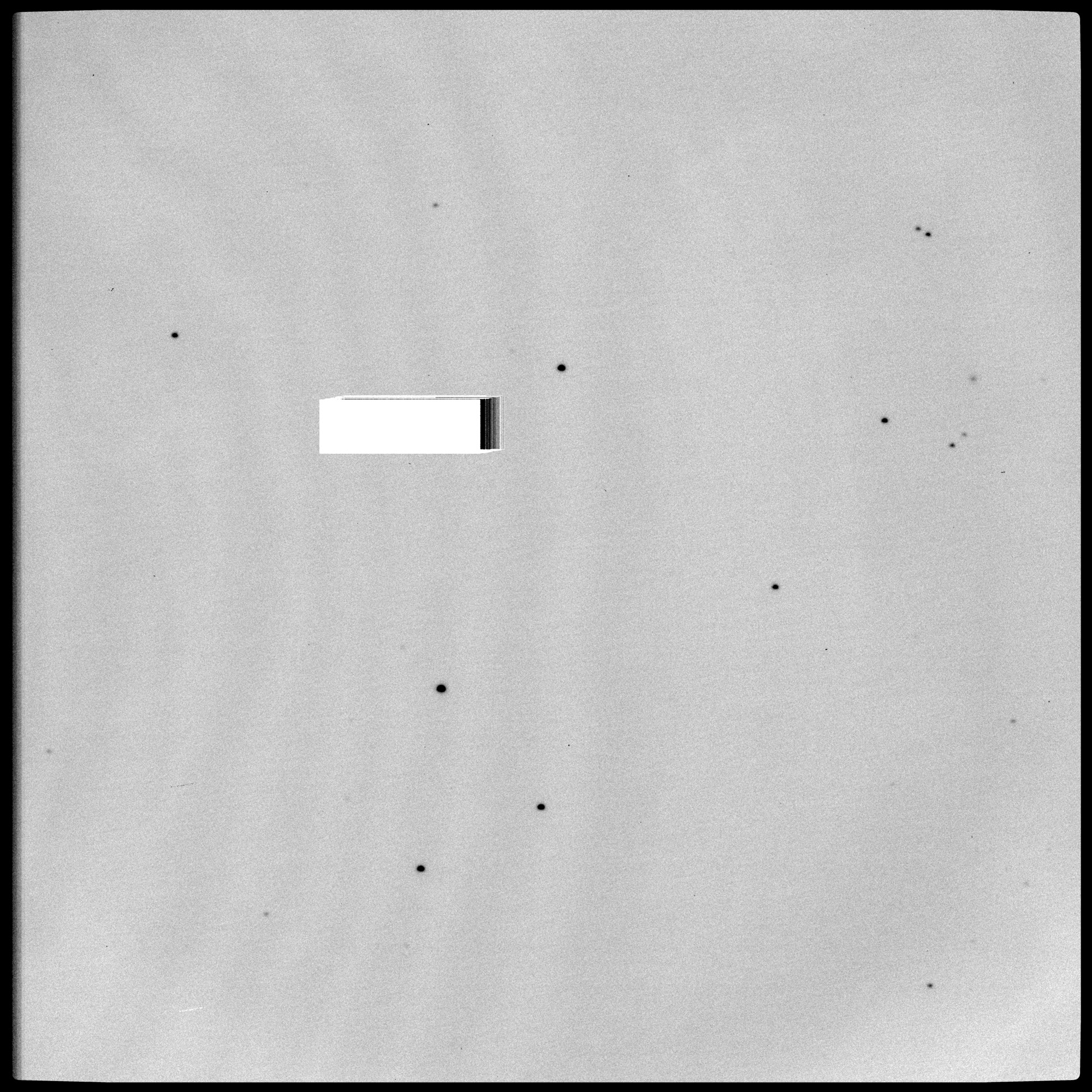}
	\caption{The superimposed reference star image. }
	\label{fig:fig5}
\end{figure}

\begin{figure}[b]
	\centering
	\includegraphics[scale=0.25]{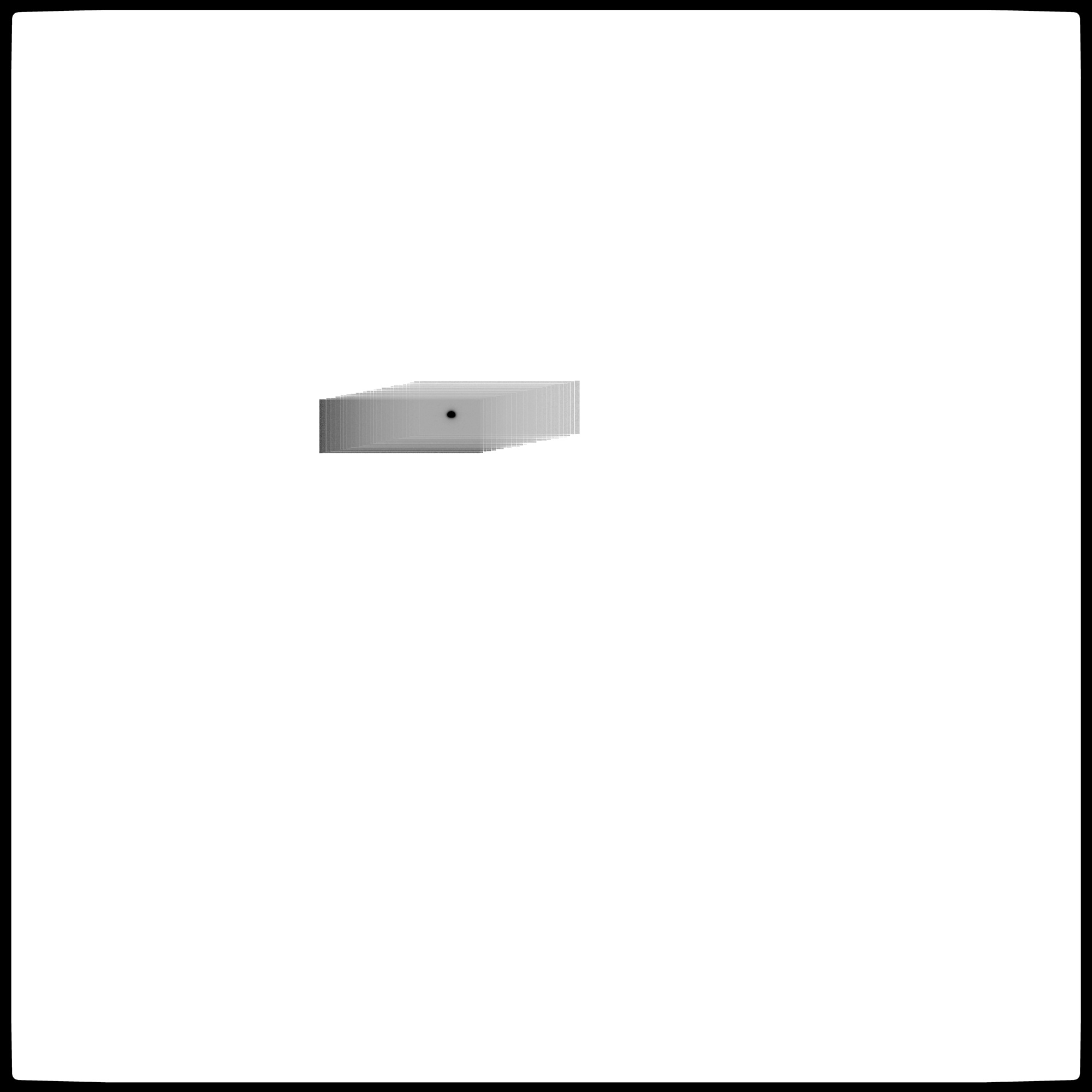}
	\caption{The superimposed asteroid image. }
	\label{fig:fig6}
\end{figure}

Step 5. We select an asteroid image as the reference image, then we align and stack other images with this image to obtain a superimposed asteroid image. The calculation method is described as follows. First, we calculate the center coordinates of the asteroid in the reference image and other images, then we obtain the translation between the reference image and other images. Finally, we perform the alignment and superposition of other asteroid images and the reference image by translation to obtain the superimposed asteroid image. The gray-scale value of the image in the superimposed region is the mean gray-scale values of the original images in the superimposed region. Figure 6 shows the processing result of this step.

Step 6. Using the above steps, we obtain the superimposed images of reference stars and asteroid at each observation time. We then fuse the superimposed images together according to the following form
\begin{equation}
{Stacking}_{Full}={Stacking}_{Star}+{Stacking}_{Asteroid},
\end{equation}
where ${Stacking}_{Full}$ is the data of the fused image, ${Stacking}_{Star}$ is the superimposed image data of reference stars, and ${Stacking}_{Asteroid}$ is the superimposed image data of asteroid. Figure 7 shows the processing result of this step. For a bright asteroid we can have another choice that we use the original image to obtain a fused image.

\begin{figure}[h]
	\centering
	\includegraphics[scale=0.25]{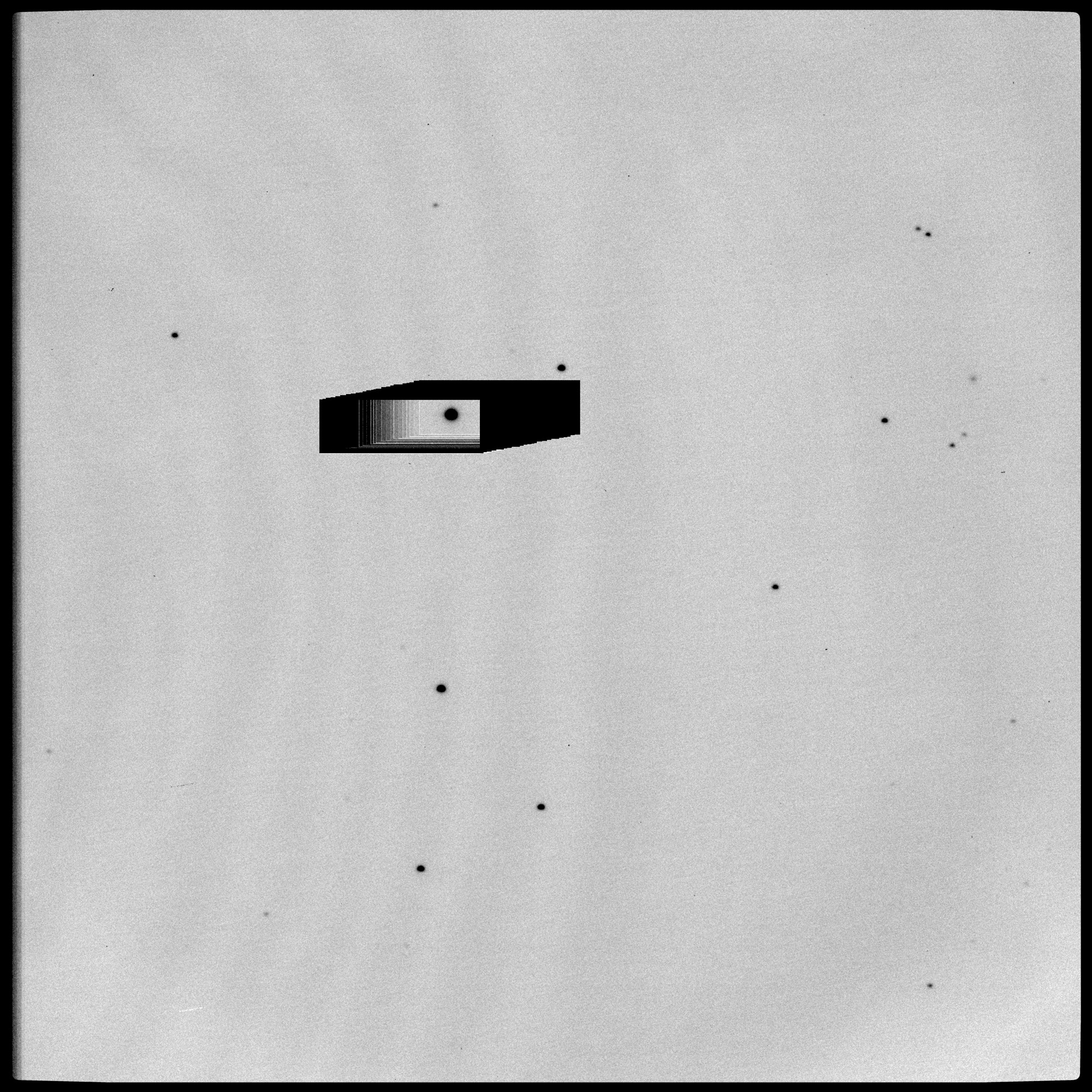}
	\caption{The fused image. }
	\label{fig:fig7}
\end{figure}

We repeat the above steps using different original image as reference image to obtain a set of the fused images, in which the number and the observation time are the same with that of the original images.

\section{Limits of applicability of the image fusion technique} \label{sec:limits}
At present, we mainly use image fusion technique for follow-up observations of NEOs, so the premise of using this technique is that the observed images of NEOs can be obtained. In order to obtain the observed images of NEOs, we need to consider the observable limiting apparent magnitude (OLAM) of NEOs at the observation station. To solve this problem,  we can use the emissive intensity and FWHN of a bright and unsaturated reference star included in the open cluster CCD image or an observed image of NEO. We can also use the minimum detectable S/N of NEOs in the CCD image as the main input parameters, and estimate the relationship between the velocity of NEOs, the exposure time for observation, and the OLAM of NEOs. The detailed solution for estimating OLAM of NEOs are presented as follows.

The apparent magnitude of objects can be estimated based on their emissive intensity. For two objects in the same field of view, and we define them as star A and B. In an observed CCD image, their apparent magnitudes satisfy the relation of
\begin{equation}
	m_A - m_B = -2.5\times lg \frac{I_A}{I_B},
\end{equation}
where $m_A$ and $m_B$ are the apparent magnitudes of star A and star B, and $I_A$ and $I_B$ are the emissive intensities of star A and star B. The S/N of a star in the observed image is given by
\begin{equation}
	SNR=\frac{I}{\sigma_{Bgd}\times \sqrt{Num}},
\end{equation}
where $I$ is the emissive intensity of the star, $\sigma_{Bgd}$ is the standard deviation of the background, and $Num$ is the number of pixels for calculating the intensity of the star. The selected area for calculating the intensity of the star is based on the effective radius $r$ of the star, then $Num =\pi \times r^2$, and the calculation of stellar intensity needs to deduct the background.

Assuming that the minimal detectable S/N for the target is $S/N_L$, the corresponding minimal intensity of the target is $I_L$. The relationship between $S/N_L$ and $I_L$ is given by

\begin{equation}
	I_L=\sigma_{Bgd}\times \sqrt{Num}\times SNR_L.
\end{equation}

We can take a CCD image containing background stars at the observation night, in which the exposure time is $T_R$. We establish the mapping relationship between the captured image and the star catalog by crossmatching, and select a bright and unsaturated star as reference star. Where its apparent magnitude is $m_R$ from the star catalog, its intensity is calculated as $I_R$ from the observed CCD image, its FWHM is $FWHM_R$, and the radius of this star is $r$.

In a single CCD image with exposure time $T_R$, when the moving distance of a object is equal to $FWHM_R$, its fastest moving speed corresponding to OLAM is defined as $V_R$. They satisfy the relation of
\begin{equation}
	V_R=\frac{FWHM_R\times K}{T_R},
\end{equation}
where $K$ is the angular extent per pixel.

For an object, we define its OLAM as $m_O$, the equivalent exposure time for observing as $T_O$, and its velocity as $V_O$. If its position is not change in the field of view, we know that $m_O$ is positively associated with $\frac{T_O}{T_R}$, implying the longer the exposure time, the higher the detection limit. If the object is moving in the field of view, we know that $m_O$ is positively associated with $\frac{V_R}{V_O}$, indicating the faster the moving object, the lower the detection limit. Based on above information and Equations $(11)\sim(14)$, we obtain the relationship between $m_O$, the velocity $V_O$ and the equivalent exposure time $T_O$ for a moving object given by

\begin{equation}
	m_O=m_R+2.5\times lg(\frac{I_R}{I_L}\times \frac{T_O}{T_R}\times \frac{V_R}{V_O}),
\end{equation}
which can be used to estimate the OLAM of a moving object based on the observed images. Then, there are two observation modes to relate with the velocity $m_O$ of a moving object. First mode: the target is a slowly moving object, e.g., $V_O\leq V_R$, this observation mode only needs to take a single image with exposure time $T_O$ to include the target and reference stars. Second mode: the target is a quickly moving object, e.g., $V_O > V_R$, this observation mode needs to take multiple images with total exposure time $T_O$, and using ``shift-and-stack'' method for avoiding streaked image, in which the exposure time of single image is given by
\begin{equation}
	T=\frac{FWHM_R \times K}{V_O},
\end{equation}
where $T$ is the exposure time of a individual image, and the number of images for stacking is $\frac{T_O}{T}$ . The intensity $I_S$ of observed moving target in single image is given by
\begin{equation}
	I_S= I_L \times\frac{T}{T_O},
\end{equation}
which is restricted by the readout noise of CCD Detector.

We can develop the observation strategy to ensure that we can get the observed images of NEOs based on above information, then we can use the image fusion technique to do astrometry. In addition, for the objects with poorly defined orbit, the image fusion technique will be limited duo to that the observed images of the objects are not easily obtained. 

\section{Results and discussion} \label{sec:result}
We have observed near-Earth asteroid Eros, during 2019. The observations were made with the 1 m optical telescope of Yunnan Observatory. 26 CCD observations were taken on 2019 June 2. The main parameters of the telescope and CCD detector are shown in Table 1. A set of the fused images on Eros are obtained using the methods presented in the paper. We use the interactive software tool Astrometrica to do the astrometric solution, adopt the point-spread function (PSF) model method based on Gaussian curve as the centering method, and choose the linear fit plate model to calibrate CCD field. The theoretical positions of Eros are retrieved from the ``Institut de mécanique céleste et de calcul des éphémérides'' (IMCCE).

The reference stars are from the Gaia DR2 star catalog with 1.7 billion star positions, and their median uncertainties in parallax and position at the reference epoch J2015.5 are about 0.04mas for bright $(G\textless14mag)$ sources, 0.1mas at \textit{G} = 17 mag, and 0.7 mas at \textit{G} = 20 mag \citep{gaia2018gaia, mignard2018gaia}. For descriptions of the measurement processes, please refer to Qiao et al \citep{qiao2008astrometric, qiao2011ccd}. We have obtained other two image sets for comparison with original image set and fused image set, and redefined the four image sets as ``SS+UA,'' ``US+SA,'' ``US+UA,'' ``SS+SA,'' where the symbols ``US,'' ``UA,'' ``SS,'' ``SA'' mean the original star image set, original asteroid image set, superimposed star image set, and superimposed asteroid image set, respectively. The astrometric results are obtained from these four image sets respectively. The (O - C)(the difference between our astrometry and the ephemeris) residuals of Eros in R.A. and decl. based on different image sets are shown in Figure 8, and the statistics on the (O - C) residuals are shown in Table 2.

\begin{figure}[ht]
	\centering
	\subfigure{\includegraphics[scale=0.31]{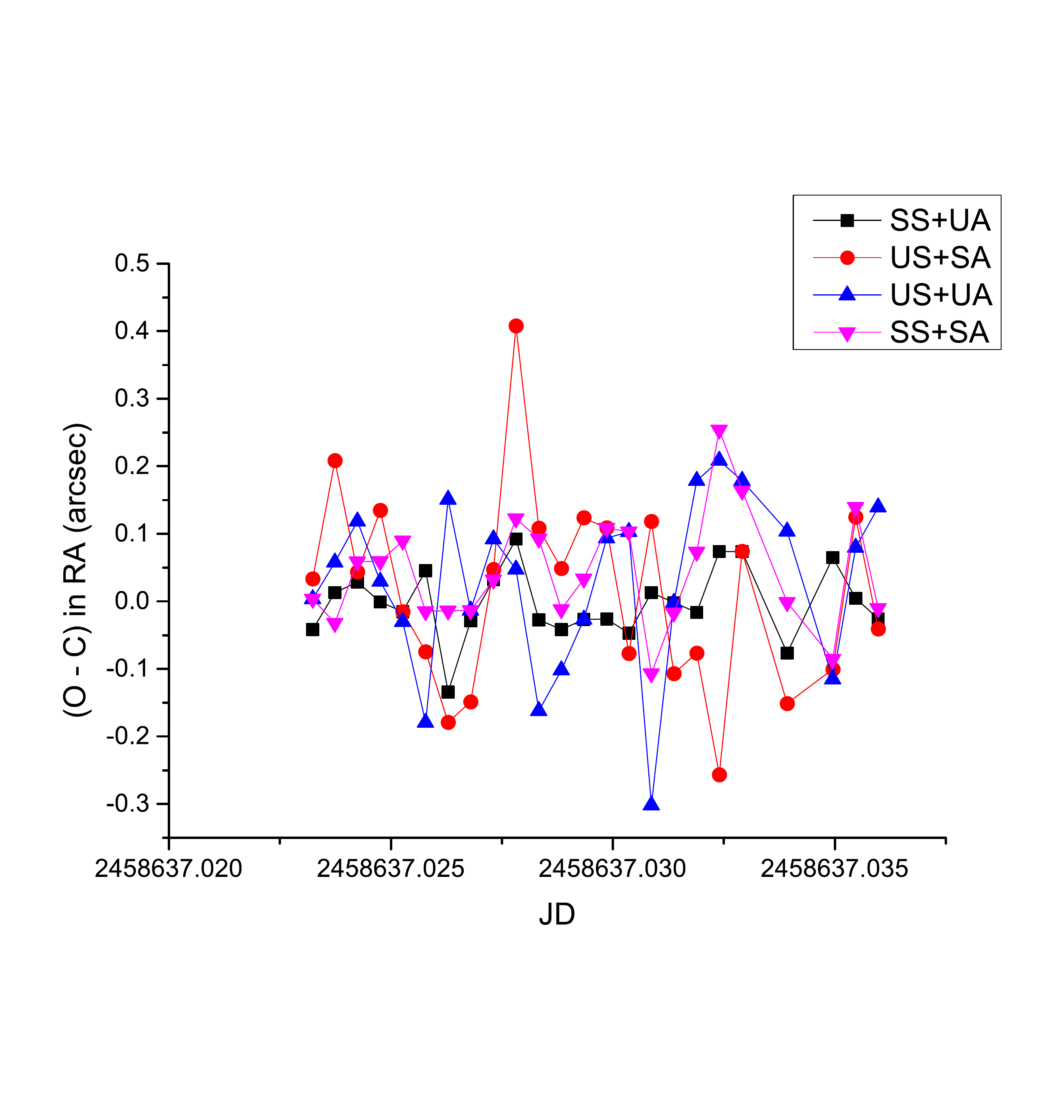}}
	\subfigure{\includegraphics[scale=0.31]{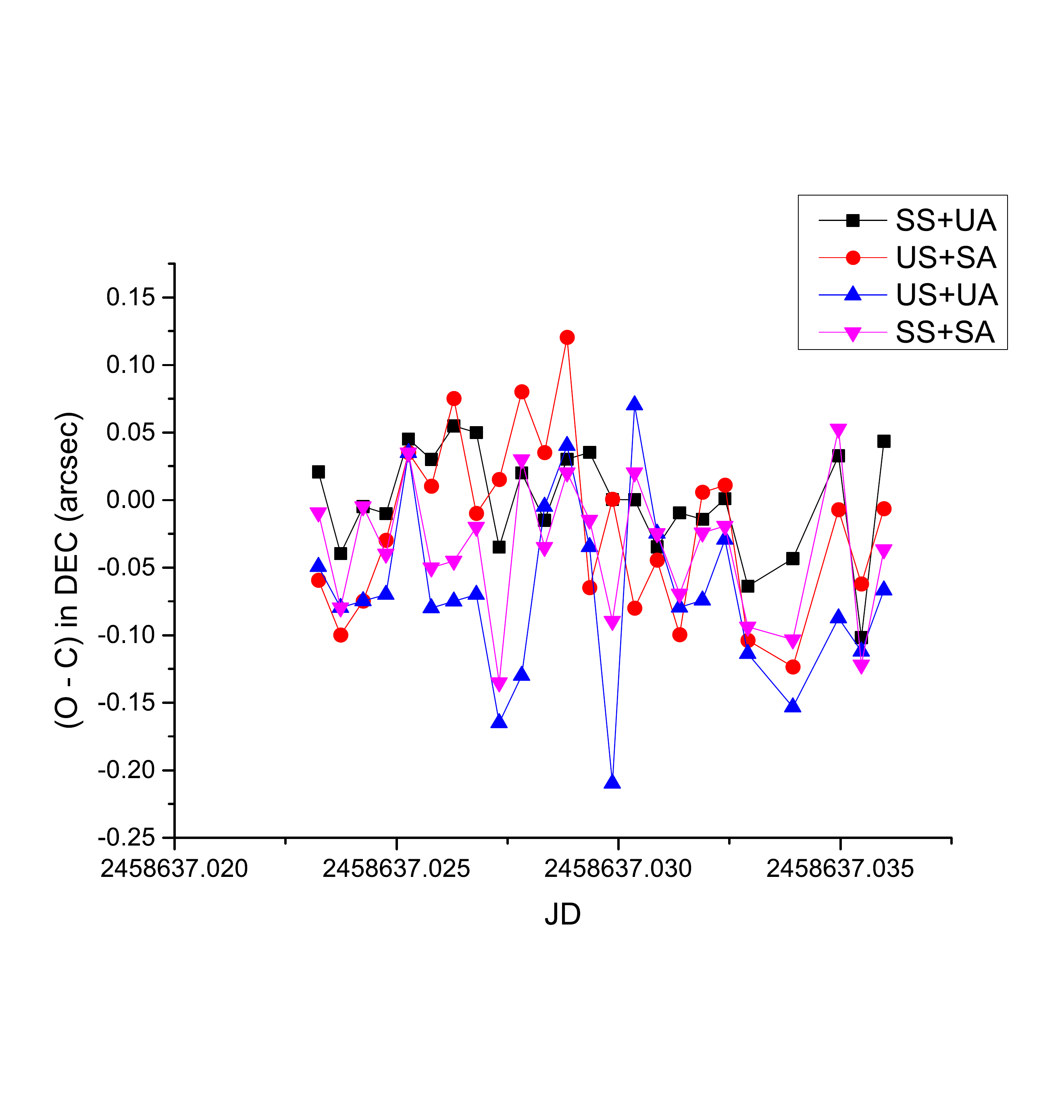}}
	\subfigure{\includegraphics[scale=0.31]{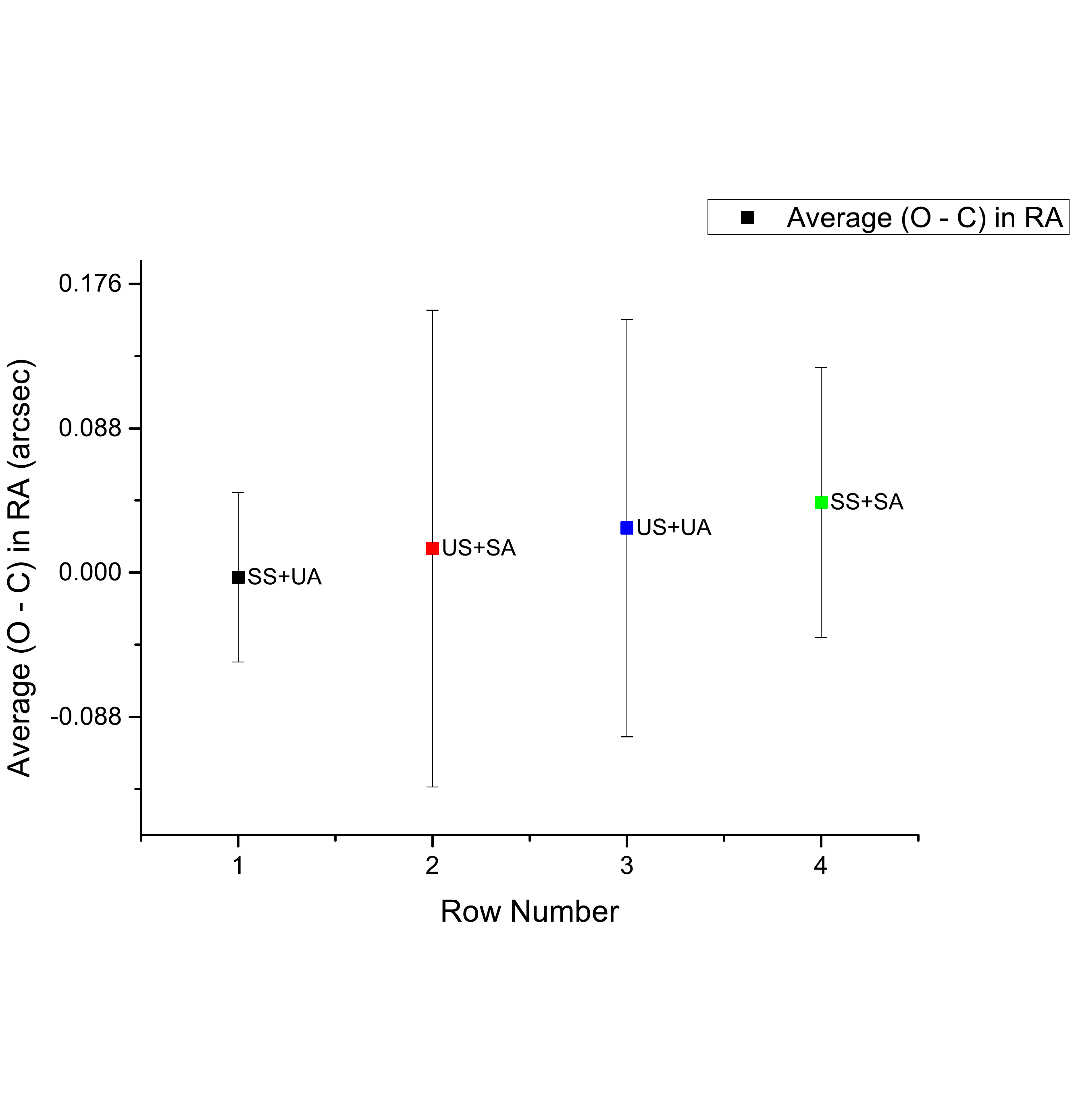}}
	\subfigure{\includegraphics[scale=0.31]{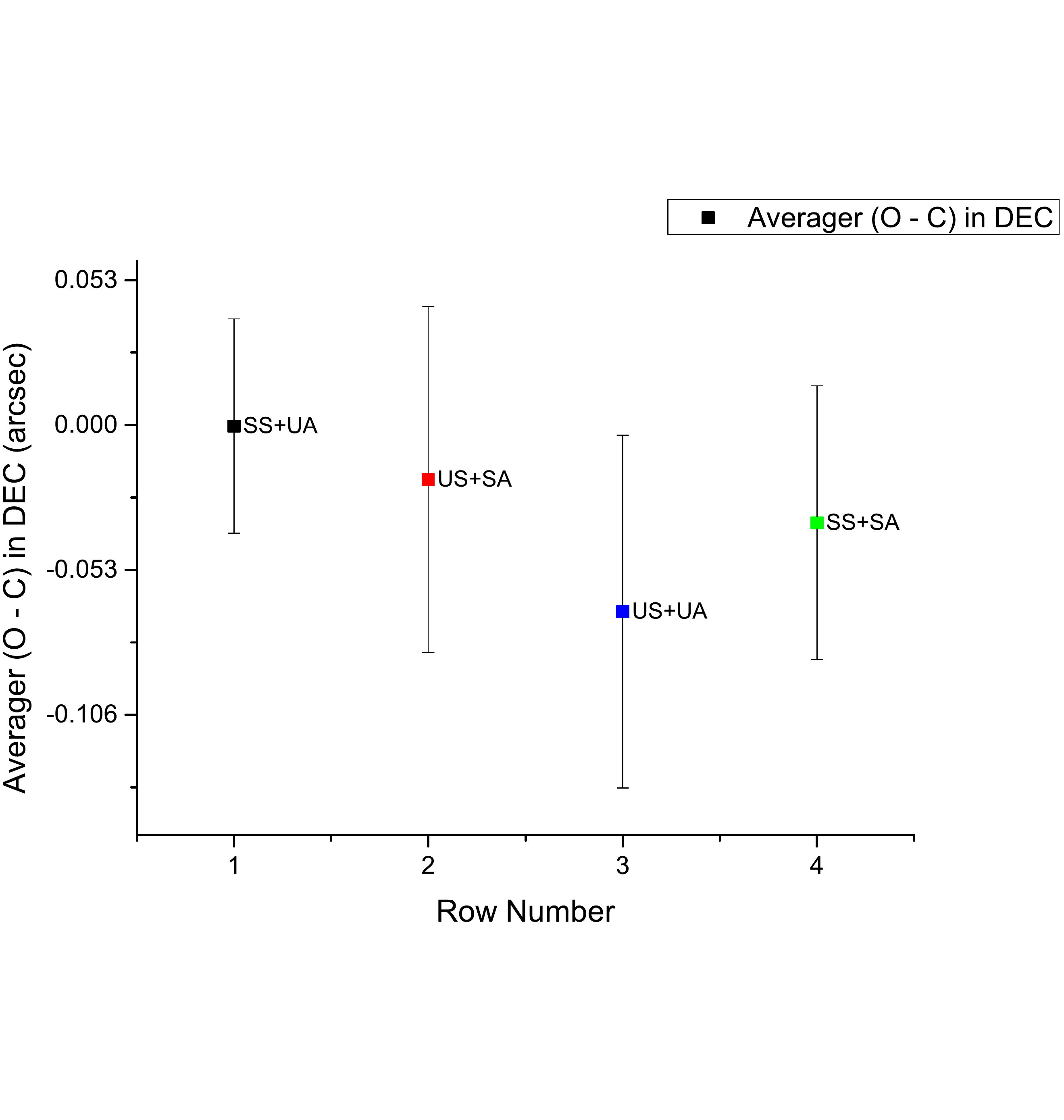}}
	\caption{The (O - C) residuals of Eros with different fuse mode. This figure shows the (O - C) residuals of Eros with different fuse mode and gives the average (O - C) residuals with the error bar.}
	\label{fig:fig8}
\end{figure}
\begin{deluxetable*}{ccccccccc}
	\tablenum{2}
	\centering
	\tablecaption{Statistics of (O - C) residuals for Eros with different fuse mode.\label{tab:result}}
	\tablewidth{5pt}
	\tablehead{
		\colhead{Eros} & \colhead{Mode} &\colhead{R.A.} & \colhead{} & \colhead{Decl.} & \colhead{} \\
		\colhead{}& \colhead{}&\colhead{(O - C)} & \colhead{SD} & \colhead{(O - C)} &\colhead{SD}}
	\startdata
	20190622 & SS+UA &-0.003 & 0.052 &  0.000 & 0.039\\
	20190622 & US+SA & 0.015 & 0.145 & -0.020 & 0.063\\
	20190622 & US+UA & 0.027 & 0.127 & -0.068 & 0.065\\
	20190622 & SS+SA & 0.043 & 0.082 & -0.035 & 0.050
	\enddata
	\tablecomments{In this table, ``Mode'' means the results are obtained using different image sets for image fuse,  ``(O - C)'' means the difference between our astrometry and the ephemeris, and ``SD'' means the positional standard deviation.}
\end{deluxetable*}

In Figure 8 and Table 2, we compare the (O - C) residuals of Eros by using different data sets. Compared with the IMCCE ephemeris,
 we obtain that the optimal mean values of (O - C) in R.A. and decl. are $-0.003''$ and $-0.000''$ based on ``SS+UA'' image set compared to $0.027''$ and $-0.068''$ based on ``US+UA''(original image set) image set.
The optimal dispersions of our observations are estimated to be about $0.052''$ and $0.039''$ in R.A. and decl. also using ``SS+UA'' image set compared to $0.127''$ and $0.065''$ using original image set. It is shown that the variances of R.A. and decl. are decreased using ``SS+UA'' image set. The precision and accuracy of astrometric observation are improved obviously. However, the results from ``US+SA'' image set are close to and even worse than those of original image set, so we analyze the FWHM and Flatness of star and Eros to find the possible causes for this as follows, in which the FWHM represents the atmospheric seeing of observation and the Flatness shows the ellipticity of star and asteroid image.

Good atmospheric seeing will increase the quality of asteroids and background stars to further affect the accuracy and precision of astrometry. We have gathered the statistics about FWHM and Flatness based on the reference star shown in Figure 3 from the fused image and the original image sets. The statistical properties are presented in Figure 9 and Table 3, showing that the variation of FWHM and Flatness for star in fused image set is smaller than that in original image set. It is indicated that the image fuse processing decrease the impact from the atmosphere that could cause different apparent offsets between exposures. In Figure 10, the variation of FWHM and Flatness for asteroid in fused image set is also smaller than that of original image set, but the average FWHM of asteroid is lager than that of star in the same view field, indicating that the asteroid image has been weakly streaked to reduce the effect of image stacking processing on astrometry. Therefore, we should reconsider image fuse mode based on the information of asteroid for the bright asteroid Eros with weak streak, and the image fuse mode with superimposed star image and original asteroid image is a better choice. We think that there are three main reasons to improve the results based on above results and analysis.

(1) The fused images contain more reference stars with high S/N for good correction by Gaia DR2 star catalog.

(2) The image quality of stars is improved for benefiting precise coordinate measurements.

(3) The effect of instrument tracking errors on astrometric observations is reduced.

\begin{figure}[ht]
	\centering
	\subfigure{\includegraphics[scale=0.25]{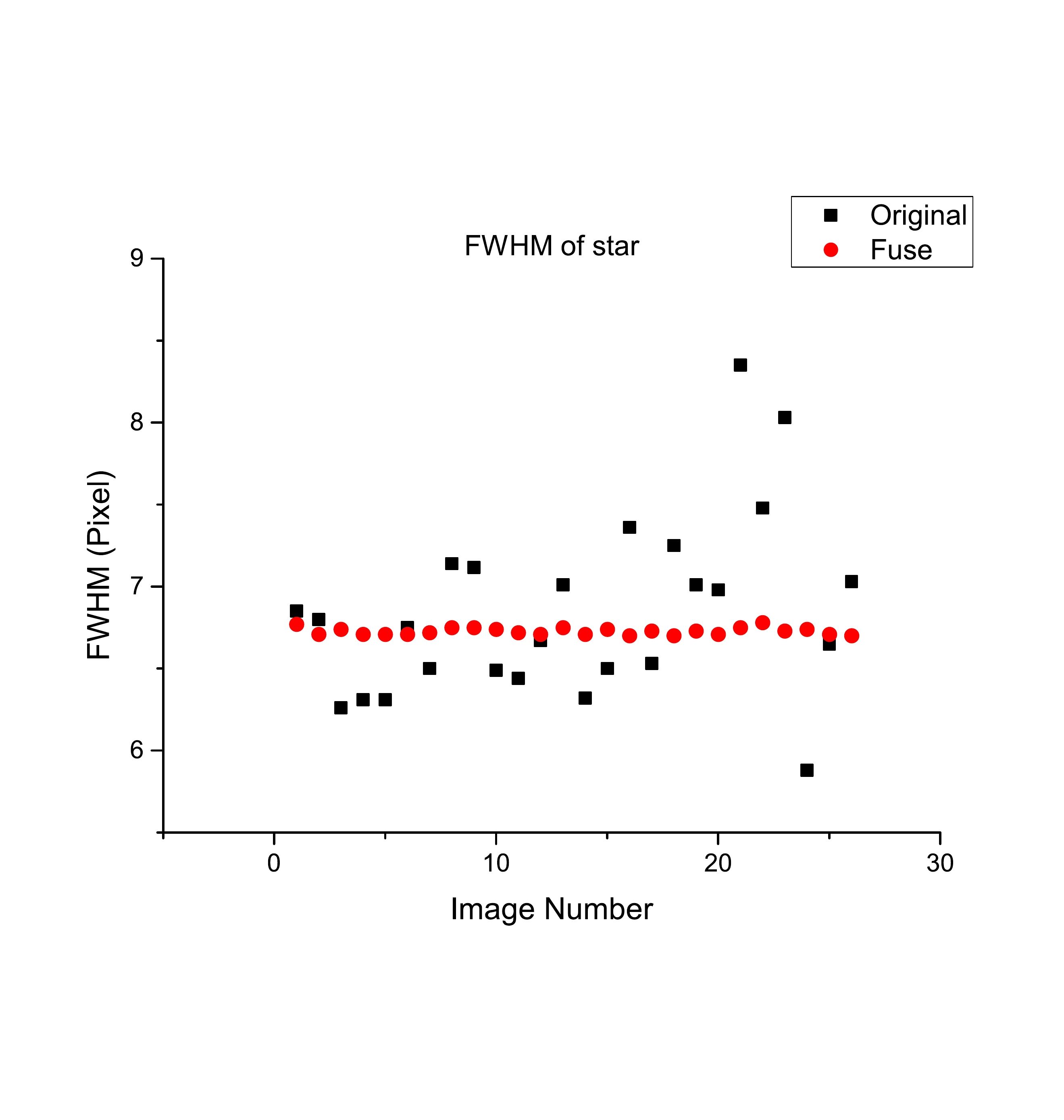}}
	\subfigure{\includegraphics[scale=0.25]{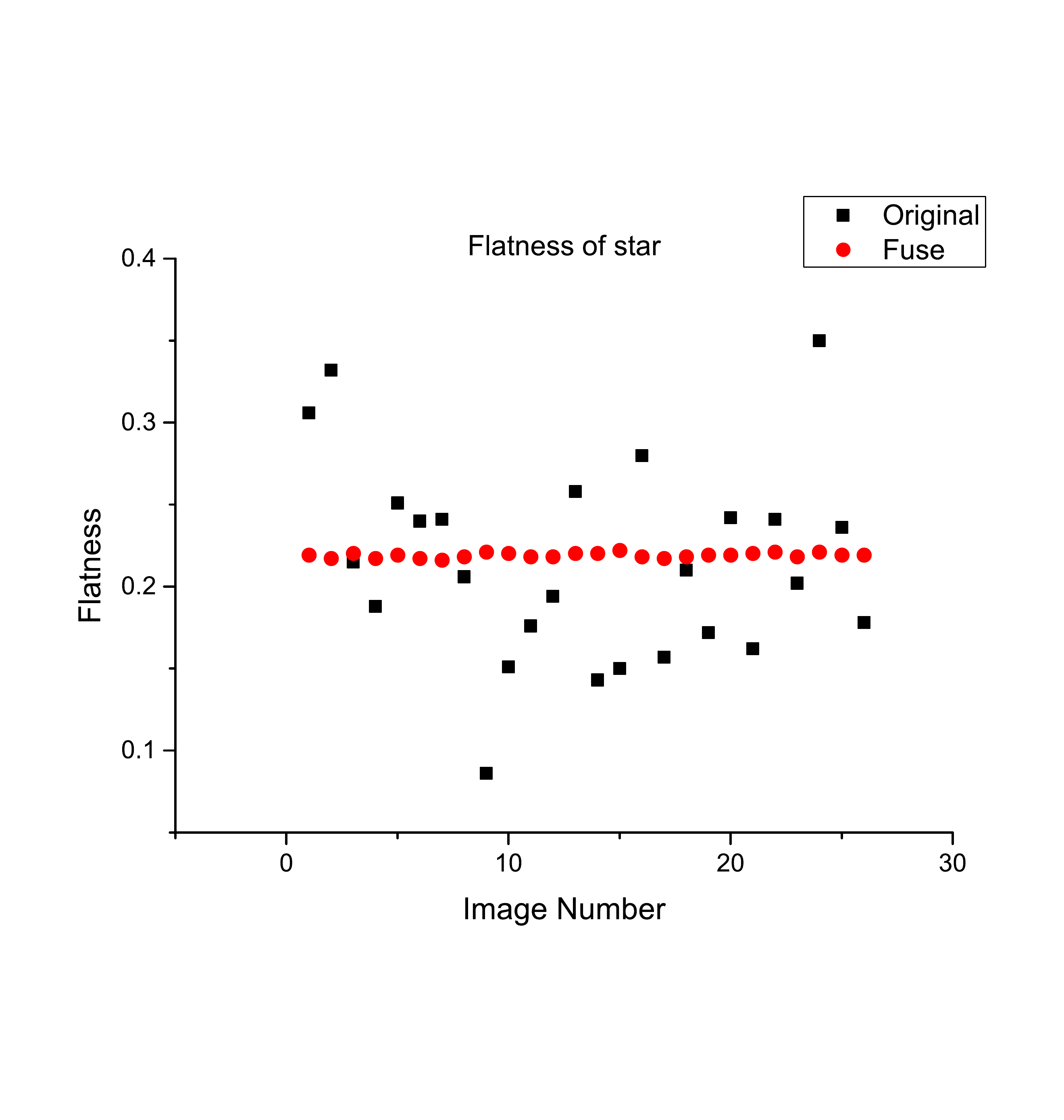}}
	\subfigure{\includegraphics[scale=0.25]{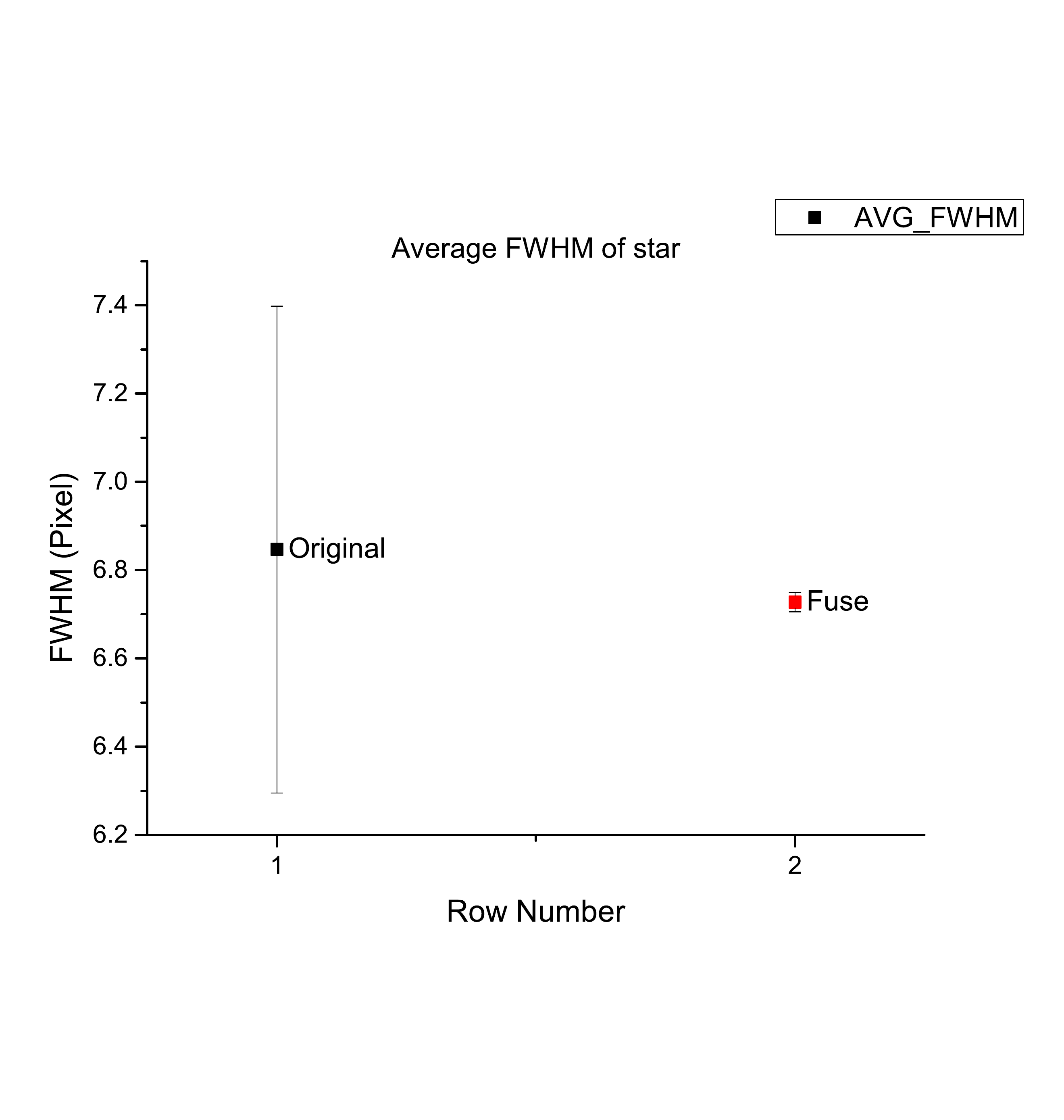}}
	\subfigure{\includegraphics[scale=0.25]{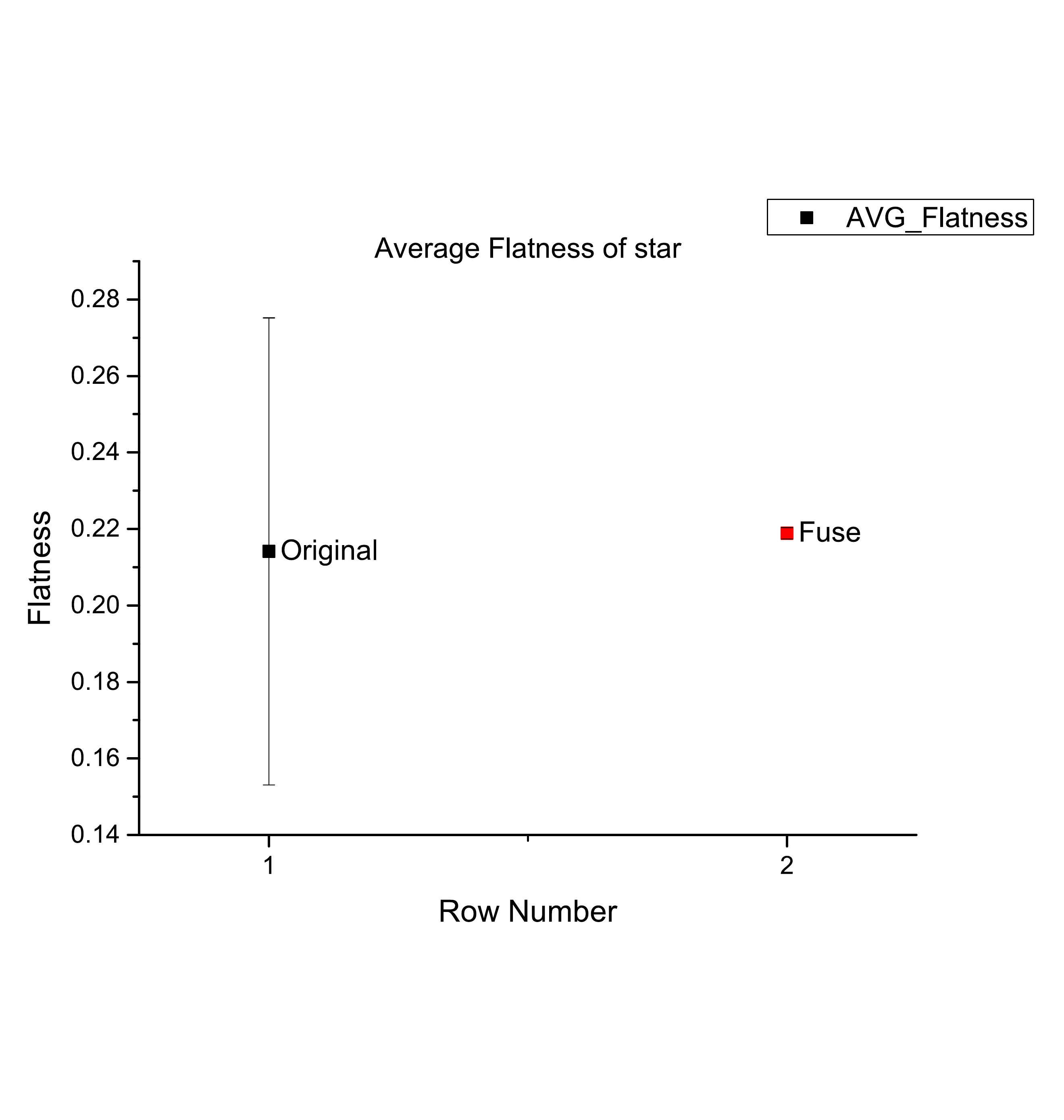}}
	\caption{The statistical properties of FWHM and Flatness for star during the period of observation.}
	\label{fig:fig9}
\end{figure}

\begin{figure}[ht]
	\centering
	\subfigure{\includegraphics[scale=0.25]{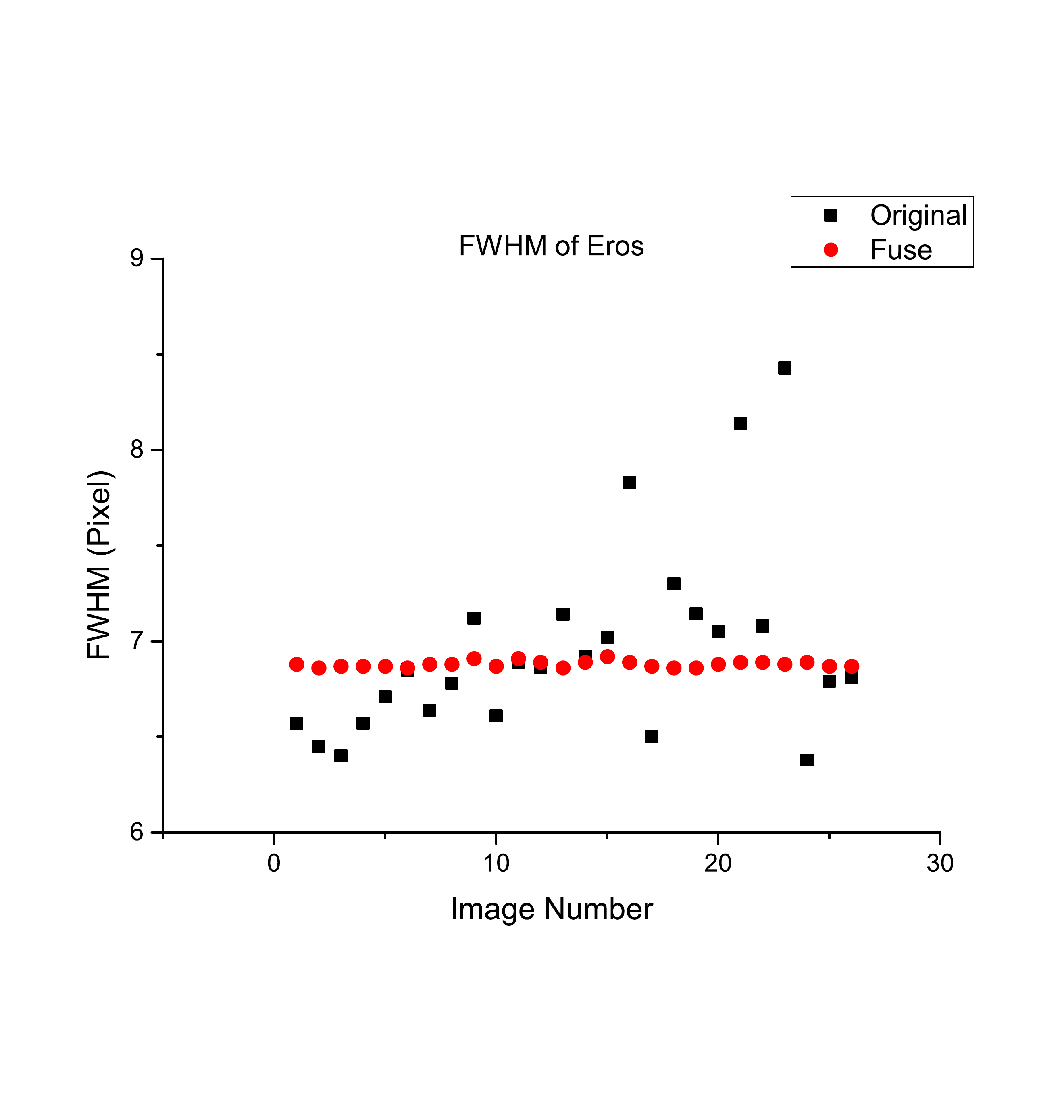}}
	\subfigure{\includegraphics[scale=0.25]{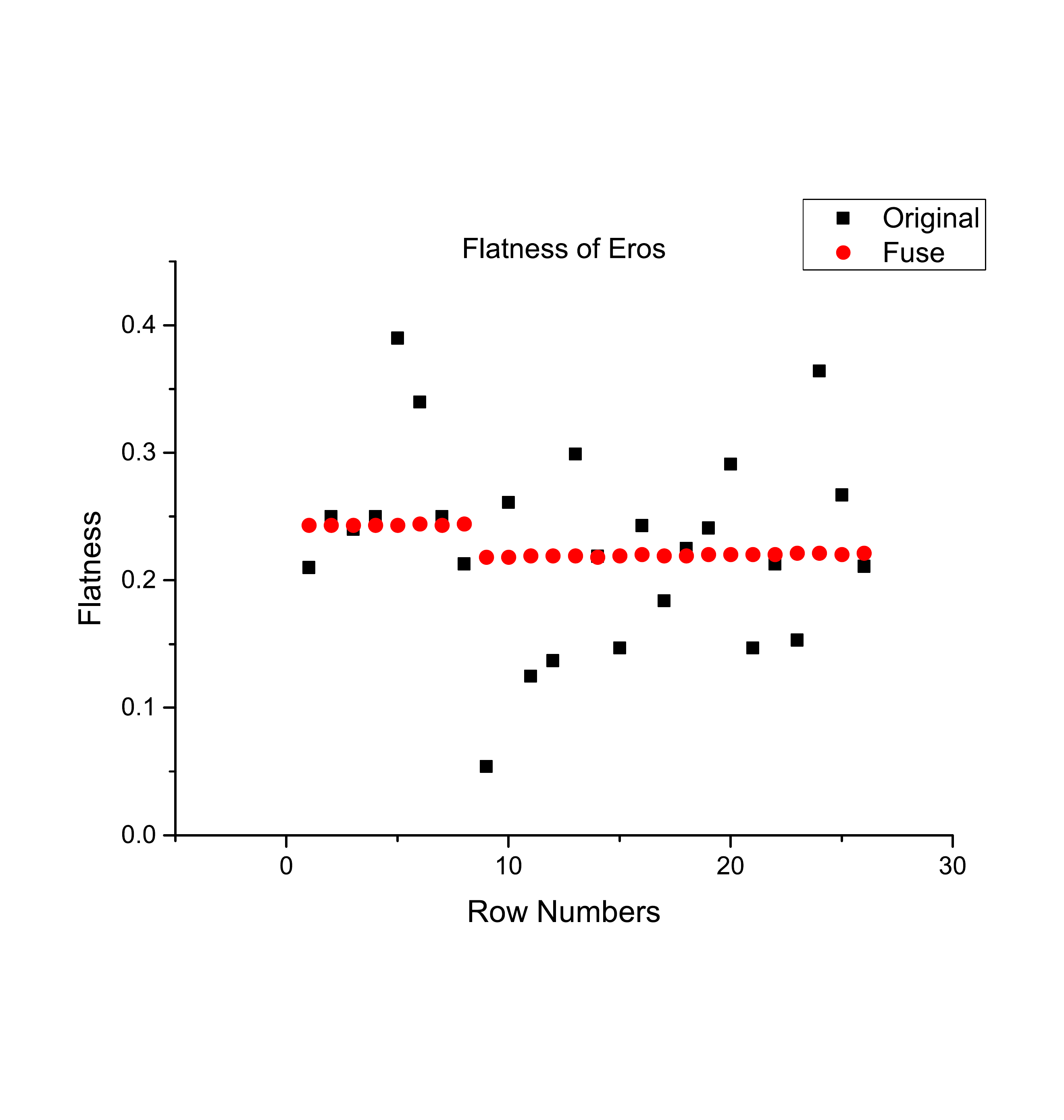}}
	\subfigure{\includegraphics[scale=0.25]{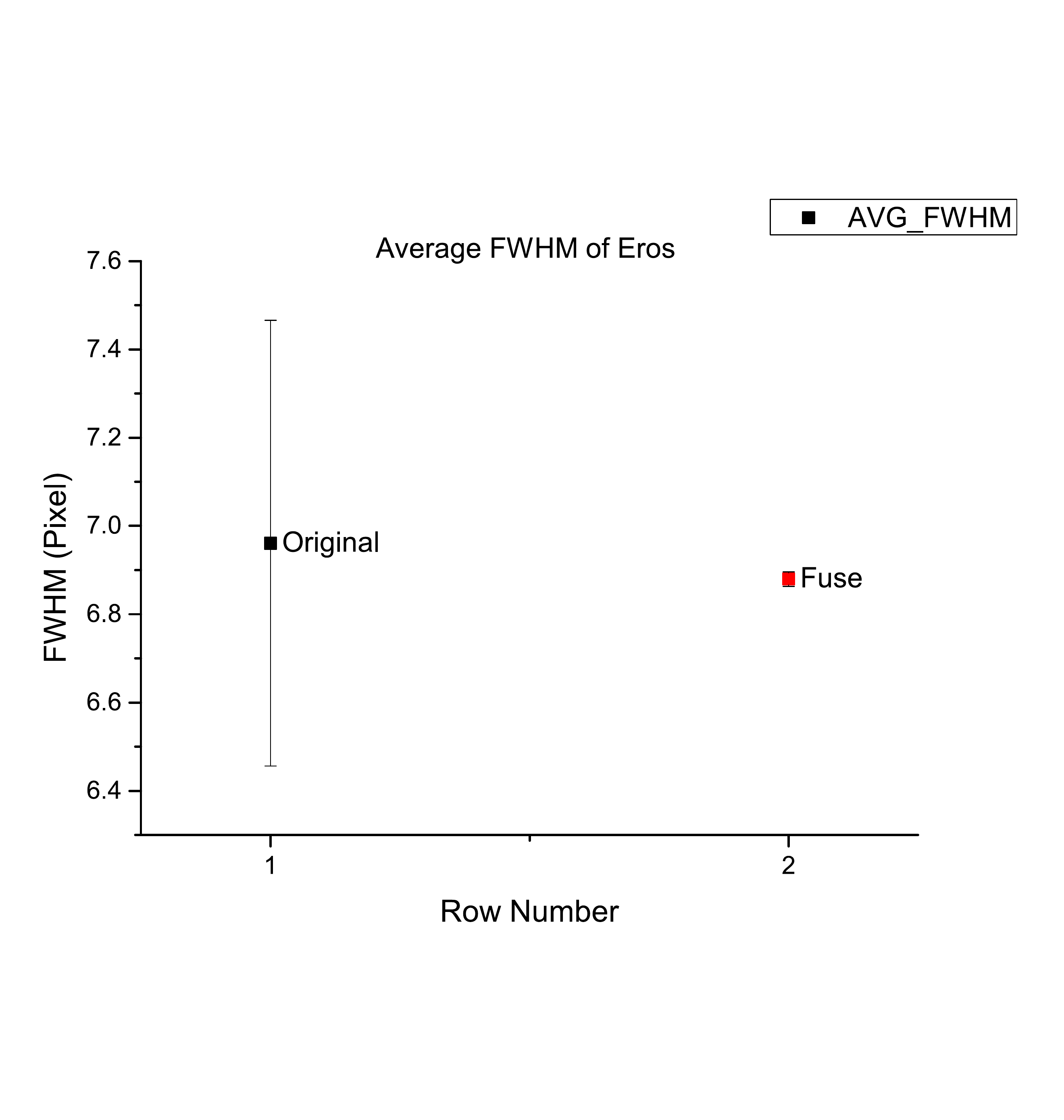}}
	\subfigure{\includegraphics[scale=0.25]{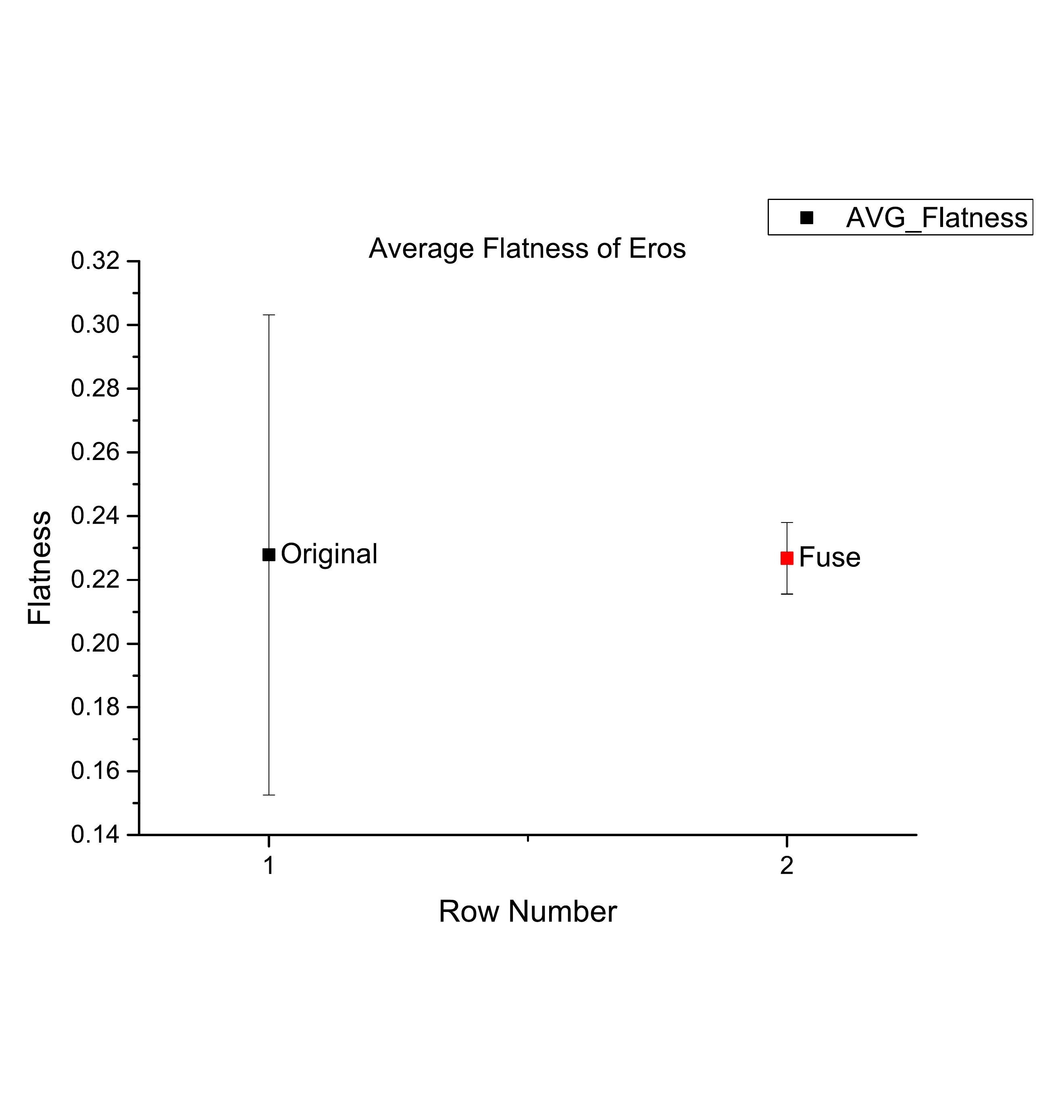}}
	\caption{The statistical properties of FWHM and Flatness for Eros during the period of observation.}
	\label{fig:fig10}
\end{figure}

\begin{deluxetable*}{ccccc}
	\tablenum{3}
	\centering
	\tablecaption{Statistics of FWMH and Flatness for star and Eros in original image set and fused image set.\label{tab:statistics}}
	\tablewidth{10pt}
	\tablehead{\colhead{} & \colhead{Original} &\colhead{} & \colhead{Fuse} & \colhead{} \\
		\colhead{} & \colhead{AVG} & \colhead{SD} & \colhead{AVG} & \colhead{SD}}
	\startdata
	FWHM (Star) & 6.846 & 0.551 & 6.727 & 0.022\\
	FWHM (Eros) & 6.961 & 0.505 & 6.879 & 0.016\\
	Flatness (Star) & 0.214 & 0.061 & 0.218 & 0.001\\
	Flatness (Eros) & 0.228 & 0.075 & 0.227 & 0.011\\
	\enddata
	\tablecomments{In this table, ``Original'' means the results are obtained using original image set, ``Fuse'' means the results are obtained using fused image set, ``AVG'' means the mean value, and ``SD'' means the positional standard deviation.}
\end{deluxetable*}

Furthermore, in the paper we use postprocessing to realize the astrometry of open cluster M23 and the near-Earth asteroid Eros, and do not consider time consumption. In future work, the processing speed should be considered. To derive more accurate astrometric results for fast moving objects, we should use a precise timing system during observation, and consider the geometric distortion of CCD images in astrometry processing \citep{anderson2002improved, peng2012convenient, wang2019distortion}. In addition, some astrometric effects, such as the precision premium \citep{lin2019characterization}), the effect of solar phase angle \citep{lindegren1977meridian}, and the differential color refraction \citep{2020Using}, should be considered carefully. 

\section{Conclusions} \label{sec:Conc}
	
In the paper, we have demonstrated the efficacy of image stacking technique on 1 m optical telescope in Yunnan Observatory in improving the S/N and astrometric results, based on the astrometry of the open cluster M23. Then we present a method of image fusion technique to carry out astrometry in postprocessing, and introduce the information about the limits of applicability of this technique. The image fusion technique is used to process the data of the near-Earth asteroid Eros observed by the 1 m optical telescope in Yunnan Observatory. The astrometric results derived from different image sets of Eros show that our method can improve the accuracy and precision of astrometry; but the image fuse mode should be carefully considered based on the information of moving objects, especially for bright moving targets the original target image could be a better choice for obtaining a fused image set.

\begin{acknowledgments}
We acknowledge the support from the staff at the 1 m telescope administered in Yunnan Observatories. This research work is financially supported by the National Natural Science Foundation of China (grant Nos. 11503083, 12173085). This work has used the data from the European Space Agency (ESA) mission Gaia (\break\url{https://www.cosmos.esa.int/gaia}), processed by the Gaia Data Processing and Analysis Consortium(DPAC; \break\url{https://www.cosmos.esa.int/web/gaia/dpac/consortium}). Funding for the DPAC has been provided by national institutions, in particular the institutions participating in the Gaia Multilateral Agreement.
\end{acknowledgments}

\end{document}